\pdfoutput=1
\documentclass[sigconf,screen]{acmart}
\usepackage{xspace,balance,tabularx,multirow}
\usepackage{graphics}
\usepackage{flushend}
\usepackage{tikz}
\usepackage{pgfplots}
\pgfplotsset{compat=1.16}
\usetikzlibrary{patterns}
\usepackage{subfig}
\usepackage[ruled, vlined, linesnumbered]{algorithm2e}
\usepackage{xcolor}
\usepackage{colortbl}
\usepackage{bbold}
\SetKwComment{Comment}{$\triangleright$\ }{}
\usepackage{enumitem}
\usepackage{tablefootnote}
\usepackage{upgreek,textgreek}

\pgfplotsset{every tick label/.append style={font=\tiny}}

\AtBeginDocument{%
  \providecommand\BibTeX{{%
    \normalfont B\kern-0.5em{\scshape i\kern-0.25em b}\kern-0.8em\TeX}}}
    
\copyrightyear{2022} 
\acmYear{2022} 
\setcopyright{acmcopyright}\acmConference[WWW '22]{Proceedings of the ACM Web Conference 2022}{April 25--29, 2022}{Virtual Event, Lyon, France}
\acmBooktitle{Proceedings of the ACM Web Conference 2022 (WWW '22), April 25--29, 2022, Virtual Event, Lyon, France}
\acmPrice{15.00}
\acmDOI{10.1145/3485447.3511959}
\acmISBN{978-1-4503-9096-5/22/04}

\makeatletter
\newcommand*\bigcdot{\mathpalette\bigcdot@{.5}}
\newcommand*\bigcdot@[2]{\mathbin{\vcenter{\hbox{\scalebox{#2}{$\m@th#1\bullet$}}}}}
\makeatother

\def\header{\vspace{1mm} \noindent}

\newcommand{\ie}{{\it i.e.},\xspace}
\newcommand{\eg}{{\it e.g.},\xspace}

\newcommand{\MM}{\mathbf{M}\xspace}
\newcommand{\PM}{\mathbf{P}\xspace}
\newcommand{\UM}{\mathbf{U}\xspace}
\newcommand{\VM}{\mathbf{V}\xspace}

\newcommand{\residue}{\mathbf{r}}
\newcommand{\appr}{{\pi}^{\prime}}
\newcommand{\ppr}{{\pi}}
\newcommand{\myppr}{{\beta}}
\newcommand{\fresidue}{\overrightarrow{\residue}}
\newcommand{\frp}{{\mathbf{p}}}
\newcommand{\fppr}{\overrightarrow{\pi}}
\newcommand{\bresidue}{\overleftarrow{\residue}}

\newcommand{\bppr}{\overleftarrow{\pi}}

\newcommand{\algot}{\textsc{Approx\textrm{-}BHPP}}
\newcommand{\algo}{\algot\xspace}

\newcommand{\mcsp}{\textsc{MCSP}\xspace}
\newcommand{\pisp}{\textsc{PISP}\xspace}

\newcommand{\fwdt}{\textsc{PI\textrm{-}Push}}
\newcommand{\bwdt}{\textsc{SS\textrm{-}Push}}
\newcommand{\fwdp}{\fwdt\xspace}
\newcommand{\bwdp}{\bwdt\xspace}

\newcommand{\piter}{\textsc{PowerIteration}\xspace}
\newcommand{\mc}{\textsc{MonteCarlo}\xspace}
\newcommand{\bpush}{\textsc{SelectivePush}\xspace}

\newenvironment{customlegend}[1][]{%
    \begingroup
    \csname pgfplots@init@cleared@structures\endcsname
    \pgfplotsset{#1}%
}{%
    \csname pgfplots@createlegend\endcsname
    \endgroup
}%

\def\addlegendimage{\csname pgfplots@addlegendimage\endcsname}

\makeatletter
\newcommand\footnoteref[1]{\protected@xdef\@thefnmark{\ref{#1}}\@footnotemark}
\makeatother

\let\oldnl\nl
\newcommand{\nonl}{\renewcommand{\nl}{\let\nl\oldnl}}

\begin{document}

\title{Efficient and Effective Similarity Search over Bipartite Graphs}

\author{Renchi Yang}
\email{renchi@hkbu.edu.hk}
\authornote{This work was done while at National University of Singapore.}
\affiliation{%
  \institution{Hong Kong Baptist University}
  \country{}
}


\begin{abstract}
Similarity search over a bipartite graph aims to retrieve from the graph the nodes that are similar to each other, which finds applications in various fields such as online advertising, recommender systems etc.
Existing similarity measures either (i) overlook the unique properties of bipartite graphs, or (ii) fail to capture high-order information between nodes accurately, leading to suboptimal result quality. Recently, {\em Hidden Personalized PageRank} (HPP) is applied to this problem and found to be more effective compared with prior similarity measures. However, existing solutions for HPP computation incur significant computational costs, rendering it inefficient especially on large graphs.

In this paper, we first identify an inherent drawback of HPP and overcome it by proposing bidirectional HPP (BHPP). Then, we formulate similarity search over bipartite graphs as the problem of approximate BHPP computation, and present an efficient solution \algo. Specifically, \algo offers rigorous theoretical accuracy guarantees with optimal computational complexity by combining deterministic graph traversal with matrix operations in an optimized and non-trivial way. Moreover, our solution achieves significant gain in practical efficiency due to several carefully-designed optimizations. Extensive experiments, comparing BHPP against 8 existing similarity measures over 7 real bipartite graphs, demonstrate the effectiveness of BHPP on query rewriting and item recommendation. Moreover, \algo outperforms baseline solutions often by up to orders of magnitude in terms of computational time on both small and large datasets.
\end{abstract}



\maketitle

\section{Introduction}\label{sec:intro}
The bipartite graph is a ubiquitous data structure used to model the relationships between two sets of heterogeneous objects, such as query-webpage, customer-product, and author-paper. Similarity search over bipartite graphs is a fundamental task in data mining and finds numerous real-world applications in online advertising \cite{antonellis2008simrank++,anastasakos2009collaborative,dey2020p}, recommender systems \cite{koren2008factorization,herlocker2004evaluating,li2013mapreduce}, biomedical analysis \cite{desantis2011simrank,pavlopoulos2018bipartite}, and other domains \cite{pan2004automatic,salton1993approaches,sun2005neighborhood}. Given a bipartite graph $G$ with two node partitions $U$ and $V$, and a node $u$ in $U$, the goal of similarity search over $G$ is to retrieve from $U$ the nodes that are similar to $u$ based on a pre-defined similarity measure. Ideally, a favorable similarity measure can not only quantify direct and indirect interactions between nodes with the consideration of the bipartite structures, but also is computationally-friendly; in other words, the similarity measure can capture complex topological information surrounding a node cost-effectively.

In the literature, a plethora of similarity measures \cite{katz1953new,adamic2003friends,jeh2003scaling,haveliwala2002topic,kleinberg1998authoritative,rothe2014cosimrank,jaccard1912distribution,tversky1977features,vijaymeena2016survey} are introduced for general graphs or sets. These measures either overlook the special properties of bipartite graphs or fail to incorporate high-order information between nodes, and hence, result in sub-par performance in bipartite graph mining tasks. In \cite{jeh2002simrank}, Jeh and Widow propose the well-known similarity measure, \ie SimRank, for bipartite graphs, which is based on the recursive definition: two nodes are similar if they are related to similar nodes. SimRank is further enhanced in the subsequent work \cite{antonellis2008simrank++,dey2020p} by incorporating the {\it evidence} metric and {\it scale-free} property into its definition. Although such SimRank-based similarity measures obtain encouraging results, they entail tremendous computational overheads due to their recursive definitions. Besides, they produce sub-optimal similarity scores in scale-free bipartite graphs \cite{dey2020p}. Recent studies \cite{deng2009generalized,epasto2014reduce} present a promising similarity measure, HPP, which is shown to achieve high result quality in various applications of bipartite graphs. Specifically, HPP is defined as the {\it Personalized PageRank} (PPR) \cite{jeh2003scaling,haveliwala2002topic} on the graph $\widehat{G}$ constructed based on bipartite graph $G$ with only nodes in $U$. Prior methods for HPP computation require up to $O(|U|^2)$ space cost for the materialization of $\widehat{G}$, which is prohibitive for large graphs. As such, existing techniques for PPR computation on general graphs can not be applied to solve the problem efficiently, posing a great technical challenge. In addition, HPP value $\pi(u,u_i)$ is {\em biased} as it describes the relevance of $u_i$ from the perspective of source node $u$ regardless of the perspective of target node $u_i$. To address this problem, we propose to measure the similarity between nodes $u$ and $u_i$ via their bidirectional HPP (BHPP), namely, $\pi(u,u_i)+\pi(u_i,u)$, which raises additional challenges in efficiency.

In this paper, we present an in-depth study on BHPP computation, and make the following contributions.
First, we formalize the similarity search over bipartite graphs as the problem of approximate BHPP query with absolute error guarantees, and pinpoint none of the existing techniques for PPR computations could be trivially applied to solve the problem efficiently. Moreover, we propose \algo, which takes as input a query node $u$ and an absolute error threshold $\epsilon$, and returns an approximate BHPP value $\beta^{\prime}(u,u_i)$ with at most $\epsilon$ absolute error for each node $u_i$ in $U$, with a near linear time complexity of $O(|E|\cdot \log{\frac{1}{\epsilon}})$, where $|E|$ represents the number of edges in the input graph. Last but not least, we conduct extensive experiments to evaluate the effectiveness of BHPP on two important graph mining tasks, and the query efficiency of \algo, using 7 real datasets. Our experimental results reveal that BHPP achieves superior performance compared to existing similarity measures, and our proposed \algo is up to orders of magnitude faster than baseline solutions.

The rest of the paper is organized as follows. Related work is reviewed in Section \ref{sec:relatedwork}. Section \ref{sec:preliminary} defines the notations and problem. In Section \ref{sec:algo}, we present the algorithmic design of \algo with several efficiency techniques. Our solution and the competitors are evaluated in Section \ref{sec:exp}. Finally, Section \ref{sec:conclude} concludes the paper. Proofs of lemmas and theorems appear in Appendix \ref{sec:proof}.

\section{Related Work}\label{sec:relatedwork}
\subsection{Similarity Search over Bipartite Graphs}
In the literature \cite{katz1953new,adamic2003friends,jeh2003scaling,haveliwala2002topic,kleinberg1998authoritative,rothe2014cosimrank,shi2020realtime,yang2021fast}, a plethora of similarity measures are proposed towards similarity search on graphs. These measures are mainly designed for general graphs and overlook bipartite structures. In Ref. \cite{jeh2002simrank}, Jeh and Widom propose bipartite SimRank for similarity search over bipartite graphs. Despite its effectiveness, SimRank suffers from a high time complexity of $O(n^4)$, where $n$ is the number of nodes in the graph. In subsequent work \cite{antonellis2008simrank++}, Antonellis {\it et al.} design an improved version of SimRank, referred to as SimRank++, and have shown an effective application of SimRank++ in the query rewriting problem of sponsored search. Recently, P-SimRank is introduced in Ref. \cite{dey2020p}, which optimizes SimRank and SimRank++ by incorporating the scale-free property of bipartite graphs. Unfortunately, these SimRank-based measures are all computationally expensive, especially for large graphs, and suffer from some inherent drawbacks \cite{dey2020p}.
In a recent work \cite{deng2009generalized}, Deng {\it et al.} propose HPP, extending PPR \cite{jeh2003scaling,haveliwala2002topic} to bipartite graphs in a generalized iterative framework. To enable on-the-fly similarity search on massive bipartite graphs, Epasto {\it et al.} \cite{epasto2014reduce} present a MapReduce framework, which supports the real-time computation of several popular similarity measures such as neighbor intersection, Jaccard's coefficient, Katz index \cite{katz1953new}, and HPP. In addition, in Ref. \cite{mei2008query}, the authors employ hitting time-based similarity measures to identify related queries from search logs. Tong {\it et al. } \cite{tong2008proximity} studies tracking node similarities in dynamic bipartite graphs.

\vspace{-1mm}
\subsection{PPR Computation}
This work is also highly related to PPR computation, as HPP is the PPR on the graph constructed from a bipartite graph. In the past years, PPR has been extensively studied in the literature, as surveyed in \cite{park2019survey}. There exists a large body of literature \cite{page1999pagerank,kamvar2003extrapolation,andersen2006local,fogaras2005towards,wang2017fora,wu2021unifying,chakrabarti2007dynamic,jung2017bepi,maehara2014computing,shin2015bear,zhu2013incremental,yoon2018tpa,lin2020index,jeh2003scaling,sarlos2006randomize,avrachenkov2007monte,liu2016powerwalk,sarkar2010fast} for single-source PPR queries. Among them, \cite{andersen2006local,kamvar2003extrapolation,zhu2013incremental,jeh2003scaling,shin2015bear,jung2017bepi,yoon2018tpa,page1999pagerank,maehara2014computing} rely on expensive matrix operations, \cite{fogaras2005towards,liu2016powerwalk,avrachenkov2007monte,sarlos2006randomize} estimate PPR values using a large number of random walks, and several recent studies \cite{wang2017fora,wu2021unifying,lin2020index} combine deterministic graph traversal with random walks for improved efficiency. Particularly, in Ref. \cite{wu2021unifying}, Wu {\it et al.} combine the {\em forward push} \cite{andersen2006local} with {\em power iterations} \cite{page1999pagerank}, which is similar in spirit to our proposed \bwdp in Section \ref{sec:bwd}. However, their method is devised for general graphs and employs push strategies that are totally different from ours. 
Another popular line of research focuses on single-source top-$k$ PPR queries \cite{wei2018topppr,wang2017fora,sun2011link,lofgren2015bidirectional,lofgren2016personalized,lofgren2014fast,maehara2014computing,zhu2013incremental,fujiwara2012fast}, which return nodes with top-$k$ highest approximate/exact PPR values.
Several studies \cite{wang2016hubppr,lofgren2015bidirectional,berkhin2006bookmark,andersen2008local,lofgren2013personalized,wang2020personalized} investigate PPR queries w.r.t. single node pairs or target nodes. 
Recent studies focus on dynamic graphs \cite{yu2013irwr,zhang2016approximate,zhu2013incremental,bahmani2010fast,ohsaka2015efficient,chakrabarti2007dynamic} and parallel/distributed settings \cite{guo2017parallel,shi2019realtime,guo2017distributed,hou2021massively,wang2019parallelizing,wang2019efficient,lin2019distributed}, which are beyond the scope of this paper.

\section{Preliminaries}\label{sec:preliminary}
\subsection{Notations}

Let $G = (U\cup V,E)$ be a bipartite graph\footnote{Following convention, we consider undirected bipartite graphs.}, where nodes can be partitioned into two disjoint sets: $U = \{u_1, u_2, \cdots, u_{|U|}\}$ with cardinality $|U|$ and $V = \{v_1, v_2,\cdots,v_{|V|}\}$ with cardinality $|V|$. Each edge $(u_i,v_j,w(u_i,v_j))$ in $E$ connects a node $u_i$ in $U$ and a node $v_j$ in $V$ with weight $w(u_i,v_j)$. Note that each edge $(u_i,v_j,w(u_i,v_j))$ is undirected; hence $w(u_i,v_j)=w(v_j,u_i)$ and $(v_j,u_i,w(v_j,u_i))\in E$. We denote by $N(u)$ (resp. $N(v)$) the set of neighbors of node $u$ (resp. $v$), and by $d(u)$ (resp. $d(v)$) the degree of node $u$ (resp. $v$).

Matrices and vectors are denoted in bold uppercase and lowercase, \eg $\MM$ and $\mathbf{x}$, respectively. We denote by $\MM(i)$ (resp. $\MM(\cdot,i)$) the $i$-th row (resp. column) vector of $\MM$, and by $\MM(i,j)$ the element at $i$-th row and $j$-th column. We use $\UM\in \mathbb{R}^{|U|\times |V|}$ (resp. $\VM\in \mathbb{R}^{|V|\times |U|}$) to represent the forward (resp. backward) transition matrix of $G$. Specifically, for each node $u_i\in U$ and each node $v_j\in V$, we have 
\begin{equation*}
\textstyle\UM(u_i,v_j)=\frac{w(u_i,v_j)}{ws(u_i)}, \textrm{and } \VM(v_j,u_i)=\frac{w(v_j,u_i)}{ws(v_j)},
\end{equation*}
where $ws(u_i)$ (resp. $ws(v_j)$) is the sum of weights of edges connecting node $u_i$ (resp. $v_j$), \ie
\begin{equation}\label{eq:d}
\textstyle ws(u_i)=\sum_{v_l\in N(u_i)}{w(u_i,v_l)}, \ ws(v_j)=\sum_{u_l\in N(v_j)}{w(v_j,u_l)}
\end{equation}
The {\em hidden transition matrix} \cite{deng2009generalized} $\PM$ for node set $U$ is defined as $\PM=\UM\cdot\VM \in \mathbb{R}^{|U|\times |U|}$, where each $(u_i,v_j)$ entry is calculated by
\begin{equation}\label{eq:pmatrix}
\textstyle \PM(u_i,u_j)=\sum_{v_l\in N(u_i)\cap N(u_j)}{\UM(u_i,v_l)\cdot\VM(v_l,u_j)}.
\end{equation}
The number of non-zero entries in $\PM$ can be up to $O(|U|^2)$ in the case where at least a node $v\in V$ has $O(|U|)$ neighbors in $U$.

\subsection{Problem Definition}\label{sec:def}

Given a bipartite graph $G$, two nodes $u,u_i$ in $U$, and restart probability $\alpha\in (0,1)$, the {\em Hidden Personalized PageRank} (HPP) \cite{deng2009generalized,epasto2014reduce} $\pi(u,u_i)$ $u_i$ w.r.t. $u$ is defined as the probability that a {\em random walk with restart} (RWR) \cite{tong2006fast} starting from $u$ would end at $u_i$. More precisely, at each step, an RWR originating from $u$ either (i) terminates at the current node $u_l$ with $\alpha$ probability, or (ii) navigates to a node $u_j$ based on the transition probability $\PM(u_l,u_j)$. Mathematically, the HPP value $\pi(u,u_i)$ is formulated as follows \cite{avrachenkov2007monte}:
\begin{align}\label{eq:hpp}
\pi(u,u_i)&\textstyle =\sum_{\ell=0}^{\infty}{\alpha(1-\alpha)^{\ell}\cdot\PM^{\ell}(u,u_i)}.
\end{align}
In essence, HPP is the {\em Personalized PageRank} (PPR) \cite{haveliwala2002topic,jeh2003scaling} on the weighted graph $\widehat{G}$ constructed based on $G$, where the node set of $\widehat{G}$ is $U$ and weights of edges are defined as $\PM(u,u_i)\ \forall{u,u_i\in U}$ \cite{deng2009generalized,epasto2014reduce}. In this paper, we refer to the PPR on $\widehat{G}$ as HPP so as to distinguish between it and the naive PPR on $G$. Recall that PPR only measures the relevance of node $u_i$ from the perspective of node $u$ and $\ppr(u,u_i)\neq \ppr(u_i,u)$ in general. Consequently, HPP is a {\em biased} similarity measure with limited effectiveness. To remedy this, we propose to model the similarity between node $u$ and $u_i$ by their {\em Bidirectional Hidden Personalized PageRank} (BHPP), {\it viz.},
\begin{equation}
\myppr(u,u_i)=\ppr(u,u_i)+\ppr(u_i,u).
\end{equation}

This paper focuses on computing {\em approximate} BHPP values. We consider {\em $\epsilon$-approximate BHPP queries}, as defined in Definition \ref{def:BHPP}.
\begin{definition}[$\epsilon$-Approximate BHPP Query]\label{def:BHPP}
Given a bipartite graph $G = (U\cup V, E)$, a query node $u\in U$ and an error threshold $\epsilon$, an $\epsilon$-approximate BHPP query returns an approximate BHPP value $\myppr^{\prime}_u(u_i)$ for each node $u_i\in U$, which satisfies
\begin{equation}\label{eq:eps-BHPP}
|\myppr(u,u_i)-\myppr^{\prime}(u,u_i)| \le \epsilon.
\end{equation}
\end{definition}

\subsection{Basic Techniques}\label{sec:basic}
\begin{algorithm}[!t]
\begin{small}
	\caption{\piter}
	\label{alg:pi}
	\KwIn{Bipartite graph $G$, initial vector $\mathbf{e}$, restart probability $\alpha$, and the number of iterations $t$.}
	\KwOut{$\boldsymbol{\pi}$.}
	$\boldsymbol{\pi}\gets \mathbf{e}$\;
    \lFor{$i\gets 1$ to $t$}{
        $\boldsymbol{\pi} \gets \mathbf{e}+(1-\alpha)\cdot(\boldsymbol{\pi}\cdot \UM)\cdot \VM $
    }
    $\boldsymbol{\pi}\gets \alpha\cdot\boldsymbol{\pi}$\;
	\Return $\boldsymbol{\pi}$\;
\end{small}
\end{algorithm}

Existing solutions \cite{epasto2014reduce,deng2009generalized} for HPP computation on a bipartite graph $G$ first construct hidden transition matrix $\PM$, and then directly apply PPR computation techniques \cite{page1999pagerank} with $\PM$. These solutions cannot deal with large graphs efficiently as they require the materialization of $\PM$, which is prohibitively expensive due to 
colossal construction time and storage space (up to $O(|U|^2)$ in the worst case). Epasto {\it et al.} proposed a MapReduce framework in Ref. \cite{epasto2014reduce} for scalable HPP computation, which relies on large amounts of computational resources. In addition, prior methods are mainly geared towards computing roughly-estimated HPP values instead of answering $\epsilon$-approximate BHPP queries. In what follows, we first introduce three fundamental techniques that are tailored to HPP computation while averting the materialization of matrix $\PM$, and then explain how to utilize them to answer $\epsilon$-approximate BHPP queries.

\subsubsection{\bf \mc \cite{fogaras2005towards}.} Recall that the HPP value $\pi(u,u_i)$ is defined as the probability that an RWR starting from $u$ terminates at $u_i$. Hence, a simple and straightforward way is to simulate a number of random walks from $u$, and then use the fraction of random walks that end at $u_i$ as an estimation of $\pi(u,u_i)$. According to \cite{fogaras2005towards}, we need to conduct $\textstyle O\left(\tfrac{2(1+\epsilon_f/3)\cdot\ln{(|U|/p_f)}}{\epsilon^2_f}\right)$ random walks in total to ensure that for every $u_i\in U$
\begin{equation}\label{eq:mc}
|\appr(u,u_i)-\ppr(u,u_i)|< \epsilon_f
\end{equation}
holds with probability at least $1-p_f$. As shown in prior work \cite{wang2017fora,lofgren2016personalized}, \mc is rather inefficient as it requires sampling a large number of random walks. Additionally, to facilitate random walk sampling on weighted graphs, \mc requires constructing {\em alias structures} \cite{walker1974new} the neighborhood of each node in the pre-processing phase, leading to an immense overhead. 

\subsubsection{\bf \piter \cite{page1999pagerank}.} \piter estimates HPP values by iteratively solving the following linear equation system \cite{page1999pagerank}, a variant of Eq. \eqref{eq:hpp}:
\begin{equation}
\boldsymbol{\pi}_u=\alpha\cdot\mathbf{e}_u + (1-\alpha)\cdot\boldsymbol{\pi}_u\cdot\PM,
\end{equation}
where $\mathbf{e}_u\in \mathbb{R}^{1\times |U|}$ is a one-hot vector which has value 1 at entry
$u$ and 0 everywhere else, $\boldsymbol{\pi}_u(u_i)=\pi(u,u_i)\ \forall{u_i\in U}$, and $\PM=\UM\cdot\VM$. Algorithm \ref{alg:pi} shows the pseudo-code of \piter for approximating $\boldsymbol{\pi}_u$ when inputting graph $G$ and $\mathbf{e}=\mathbf{e}_u$. Note that Algorithm \ref{alg:pi} eliminates the need to materialize $\PM$ by decoupling and reordering the matrix multiplication $\boldsymbol{\pi} \cdot \PM$ to $(\boldsymbol{\pi}\cdot \UM) \cdot \VM$ (Line 2), reducing the cost of the matrix-vector multiplications in each iteration to $O(|E|)$. To obtain an estimation of $\pi(u,u_i)$ with at most $\epsilon_f$ absolute error, Algorithm \ref{alg:pi} requires at least $t=\log_{\tfrac{1}{1-\alpha}}{\tfrac{1}{\epsilon_f}}-1$ \cite{berkhin2005survey} iterations of matrix multiplications, which results in total time complexity of $O(|E|\cdot\log{({1}/{\epsilon_f})})$. Given a large error threshold, \eg $\epsilon_f=0.1$, \piter still needs around $14$ iterations of matrix-vector multiplications, which is considerably expensive.

\begin{algorithm}[!t]
\begin{small}
	\caption{\bpush}
	\label{alg:bp}
	\KwIn{Bipartite graph $G$, target node $u$, restart probability $\alpha$ and an error threshold $\epsilon_b$.}
	\KwOut{$\{\bppr(u_i,u)|u_i\in U\},\bresidue_{u}(\cdot)$.}
	$\bresidue_{u}(u)\gets 1; \bresidue_{u}(x)\gets 0 \ \forall{x\in U\cup V}$ and $x\neq u$\;
	$\bppr(u_i,u)\gets 0\ \forall{u_i\in U}$\;
    \While{\it true}{
        \For{$u_i\in U$ s.t. $\bresidue_{u}(u_i)> \epsilon_b$}{
            \For{$v_j\in N(u_i)$}{
                $\bresidue_{u}(v_j) \gets \bresidue_{u}(v_j)+ (1-\alpha)\cdot\frac{w(v_j,u_i)}{ws(v_j)}\cdot \bresidue_{u}(u_i)$\;
            }
        	$\bppr(u_i,u)\gets \bppr(u_i,u)+\alpha\cdot\bresidue_{u}(u_i)$\;
        	$\bresidue_{u}(u_i) \gets 0$\;
        }
        
    	\For{$ v_i\in V$ s.t. $\bresidue_u(v_i)> 0$}{
    	    \For{$u_j\in N(v_i)$}{
        	    $\bresidue_{u}(u_j) \gets \bresidue_{u}(u_j)+\frac{w(u_j,v_i)}{ws(u_j)}\cdot \bresidue_{u}(v_i)$\;
    	    }
    	    $\bresidue_{u}(v_i)\gets 0$\;
    	}
    	\lIf{$\forall{u_i}\in U$ s.t. $\bresidue_{u}(u_i)\le \epsilon_b$}{
    	\bf{break}
    	}
    }
	\Return $\{\bppr(u_i,u)|u_i\in U\}$\;
\end{small}
\end{algorithm}

\subsubsection{\bf \bpush \cite{andersen2008local}.} Unlike \mc and \piter approaches, which return approximate HPP values w.r.t. a source node $u$, \bpush estimates HPP values to a target node $u$, \ie $\pi(u_i,u)\ \forall{u_i\in U}$. Algorithm \ref{alg:bp} illustrates the pseudo-code of \bpush. In brief, \bpush is a deterministic version of \mc, which recursively pushes {\em residues} (\ie the portion of RWRs that are not stopped yet) along edges during a graph traversal of $G$ from $u$. Initially, Algorithm \ref{alg:bp} sets residue $\bresidue_u(u)=1$ for source node $u$ and 0 for other nodes in $G$, as well as approximate HPP values $\bppr(u_i,u)=0\ \forall{u_i\in U}$ (Lines 1-2). Next, it iteratively pushes the residues of the selected nodes to their neighbors. In each iteration, given an absolute error threshold $\epsilon_b$, if there is any node $u_i\in U$ with residue $\bresidue_u(u_i)> \epsilon_b$ (Line 4), for each neighbor $v_j$ of $u_i$, \bpush increases $v_j$'s residue by $(1-\alpha)\cdot\frac{w(v_j,u_i)}{ws(v_j)}\cdot \bresidue_{u}(u_i)$, and transfers $\alpha$ portion of $\bresidue_{u}(u_i)$ to $u_i$'s approximate HPP $\bppr(u_i,u)$ (Lines 5-7). The residue $\bresidue_{u}(u_i)$ is set to $0$ when $u_i$'s neighbors are all processed (Line 8). Subsequently, Algorithm \ref{alg:bp} pushes residues of nodes in $V$ back to nodes in $U$. More specifically, if any node $v_i$ in $V$ has a non-zero residue (Line 9), for $v_i$'s each neighbor $u_j$, Algorithm \ref{alg:bp} increases $u_j$'s residue by $(1-\alpha)\cdot\frac{w(u_j,v_i)}{ws(u_j)}\cdot \bresidue_{u}(v_i)$ (Lines 10-11). After that, Algorithm \ref{alg:bp} resets $\bresidue_{u}(v_i)$ to 0 (Line 12). \bpush repeats the above procedures until residues of nodes in $U$ are all less than $\epsilon_b$ and none of the nodes in $V$ have positive residues (Line 13). Notice that Algorithm \ref{alg:bp} differs from the original \bpush for PPR computation. Particularly, in Algorithm \ref{alg:bp}, we push residues between $U$ and $V$ alternatively, so as to avoid the materialization of matrix $\PM$. Lemma \ref{lem:bwd-eq} shows a crucial property of Algorithm \ref{alg:bp}.
\begin{lemma}\label{lem:bwd-eq}
Consider any iteration in Algorithm \ref{alg:bp} (Lines 4-13). At the end of the iteration, the following equation holds.
\begin{equation}\label{eq:bwd-eq}
\textstyle \ppr(u_i,u)=\bppr(u_i,u)+\sum_{u_j\in U}{\pi(u_i,u_j)\cdot\bresidue_u(u_j)}.
\end{equation}
\end{lemma}
Since Algorithm \ref{alg:bp} terminates when the residue $\bresidue_u(u_i)\le \epsilon_b$ for each node $u_i$ in U, Lemma \ref{lem:bwd-eq} implies that each HPP value $\bppr(u_i,u)$ returned by Algorithm \ref{alg:bp} satisfies
\begin{equation}\label{eq:bwd}
|\ppr(u_i,u)-\bppr(u_i,u)|< \epsilon_b. 
\end{equation}
\begin{lemma}\label{lem:bwd-time}
Algorithm \ref{alg:bp} runs in $O\textstyle\left(\tfrac{\sum_{v_i\in V}{d(v_i)^2}}{|U|\cdot\alpha\cdot \epsilon_b}\right)$ amortized time.
\end{lemma}
Lemma \ref{lem:bwd-time} provides the amortized cost of \bpush. \bpush runs fast in practice due to its consideration of the residue at each node adaptively. However, it fails to calculate high-precision HPP values for graphs with large average degrees efficiently. In the worst case, the time complexity is $O\left(\tfrac{|E|}{\alpha\cdot \epsilon_b}\right)$, worse than the $O(|E|\cdot\log{({1}/{\epsilon_f})})$-bound of \piter.

\subsubsection{\bf Baselines and Challenges.}\label{sec:baseline}
An $\epsilon$-approximate BHPP query (see Definition \ref{def:BHPP}) of node $u$ asks for an approximation of exact BHPP value $\myppr(u,u_i)=\pi(u,u_i)+\pi(u_i,u)$ for each node $u_i$ in $U$ with at most $\epsilon$ absolute error. A straightforward way to answer the $\epsilon$-approximate BHPP query of node $u$ in a probabilistic fashion is to compute $\appr(u,u_i)$ by letting $\epsilon_f=\frac{\epsilon}{2}$ in \mc and get $\bppr(u_i,u)$ by \bpush with $\epsilon_b=\frac{\epsilon}{2}$ for each node $u_i$ in $U$. According to Eq. \eqref{eq:mc} and Eq. \eqref{eq:bwd}, the summed value $\appr(u,u_i)+\bppr(u_i,u)$ satisfies Eq. \eqref{eq:bwd-eq} with a high probability. Similarly, another approach for $\epsilon$-approximate BHPP queries is to invoke \piter and \bpush with $t=\log_{\frac{1}{1-\alpha}}{\frac{2}{\epsilon}}-1$ and $\epsilon_b=\frac{\epsilon}{2}$, respectively. However, due to the inefficiency of the \mc and \piter approaches, as well as the deficiency of \bpush, both two aforementioned solutions for $\epsilon$-approximate BHPP queries incur vast costs, especially for large graphs. Towards this end, there are three technical challenges that we need to address: 
\begin{enumerate}[leftmargin=*]
\item How to overcome the inherent drawback of \bpush and reduce its cost to $\textstyle O(|E|\cdot\log{({1}/{\epsilon_f})})$?
\item How to devise an algorithm that improves over \piter in terms of practical efficiency without degrading its theoretical guarantees?
\item How to integrate the above two algorithms in an optimized way for improved efficiency?
\end{enumerate}

\section{The \algo Algorithm}\label{sec:algo}

To circumvent these challenges, we first propose an optimized version of \bpush, called \underline{S}elective and \underline{S}equential \underline{Push} (hereafter \bwdp), in Section \ref{sec:bwd}; after that, we present \fwdp (short for \underline{P}ower \underline{I}terations-based \underline{Push}) to mitigate the efficiency issue of \piter in Section \ref{sec:fwd}, and then elaborate the integration of \bwdp and \fwdp to obtain our main proposal solution, \algo, for answering $\epsilon$-approximate BHPP queries.

\begin{algorithm}[!t]
\begin{small}
	\caption{\bwdp}
	\label{alg:bwd}
	\KwIn{Bipartite graph $G$, target node $u$, restart probability $\alpha$ and an error threshold $\epsilon_b$.}
	\KwOut{$\{\bppr(u_i,u)|u_i\in U\},\bresidue_{u}$.}
	{\nonl Lines 1-2 are the same as Lines 1-2 in Algorithm \ref{alg:bp}}\;
	\setcounter{AlgoLine}{2}
    $n_p\gets 0$\;
    \tcc{Selective pushes}
    \While{\it true}{
        {\nonl Lines 5-9 are the same as Lines 4-8 in Algorithm \ref{alg:bp}}\;
        \setcounter{AlgoLine}{9}
        $\quad n_p\gets n_p+N(u_i)$\;
        {\nonl Lines 11-14 are the same as Lines 9-12 in Algorithm \ref{alg:bp}}\;
        \setcounter{AlgoLine}{14}
        $\quad n_p\gets n_p+N(v_i)$\;
        \If{$\forall{u_i}\in U$ s.t. $\bresidue_{u}(u_i)\le \epsilon_b$ or}{
            \Return $\{\bppr(u_i,u)|u_i\in U\},\bresidue_{u}$\;
        }
        \lIf{Inequality \eqref{eq:bwd-np} holds}{
            \bf{break}
        }
    }
    \tcc{Sequential pushes}
        \While{$\exists{u_i}\in U$ s.t. $\bresidue_{u}(u_i)> \epsilon_b$ and $\sum_{u_i\in U}{\bresidue_u(u_i)}>\epsilon_b$}
        {
        {\bf for} $\exists u_i\in U$ s.t. $\bresidue_{u}(u_i)> 0$ {\bf do}\\
        {\nonl Lines 21-28 are the same as Lines 5-12 in Algorithm \ref{alg:bp}}\;
        \setcounter{AlgoLine}{28}
        }
    \Return $\{\bppr(u_i,u)|u_i\in U\},\bresidue_{u}$\;
\end{small}
\end{algorithm}

\subsection{\bwdp}\label{sec:bwd}
Before diving into the algorithmic details of \bwdp, we give a high-level idea of \bwdp. \bwdp suffers from severe efficiency issues in some cases. To explain, consider a node $v\in V$ of input bipartite graph $G$, which connects to $100$ neighbors $u_1$-$u_{100}$ in node set $U$. According to Lines 4-12 in Algorithm \ref{alg:bp}, in each iteration, $v$ first (i) receives residues from the selected neighbors, and then (ii) conducts 100 push operations to $u_1$-$u_{100}$. After a few iterations, only few neighbors of $v$ would be selected as the residues of majority nodes are slightly less than $\epsilon_b$. As a consequence, to deplete a certain amount of $v$'s residue, \bpush requires numerous iterations, each of which involves at least 100 push operations and random access to adjacent nodes. A promising option to alleviate this problem is to leverage {\em sequential} pushes in each iteration, which aggregate residues from $v$'s neighbors in one batch before pushing back to $u_1$-$u_{100}$. However, the sequential strategy performs push operations regardless of the residue at each node, leading to a large number of redundant push operations. To overcome the limitations of both strategies, we resort to combining them in a greedy and adaptive manner. Specifically, we execute selective pushes while recording its actual cost, and switch to sequential pushes once the recorded cost of the former exceeds the estimated computational cost using the latter. The switch is dynamically controlled by a carefully designed threshold. Below we present the details.

\subsubsection{\bf Details}
Algorithm \ref{alg:bwd} displays the pseudo-code of \bwdp. Given a bipartite graph $G$, a target node $u$, restart probability $\alpha$ and an absolute error threshold $\epsilon_b$ as inputs, \bwdp begins by initializing residue vector $\bresidue_u$ and approximate HPP $\bppr(u_i,u)$ for $u_i\in U$ as Lines 1-2 in Algorithm \ref{alg:bp}, and the number of performed push operations $n_p=0$ (Lines 1-3). After that, Algorithm \ref{alg:bwd} starts the iterative process of selective pushes as Lines 4-12 in Algorithm \ref{alg:bp}, during which the recorded cost $n_p$ of selective pushes is increased by $N(u_i)$ (resp. $N(v_i)$) if the neighboring nodes of $u_i\in U$ (resp.  $v_i\in V$) are accessed (Lines 5-14). Algorithm \ref{alg:bwd} terminates the iterative process and returns $\{\bppr(u_i,u)|u_i\in U\}$ with $\bresidue_u$ when every residue $\bresidue_u(u_i)$ is not greater than $\epsilon_b$ (Lines 16-17).
\begin{small}
\begin{equation}\label{eq:bwd-np}
\textstyle n_p\ge 2|E|\cdot \left(\log_{\frac{1}{1-\alpha}}{\frac{1}{\sum_{u_i\in U}{\bresidue_u(u_i)}}}\right) 
\end{equation}
\end{small}
Additionally, at Line 18, if Algorithm \ref{alg:bwd} depletes its computation budget for selective pushes (\ie Eq. \eqref{eq:bwd-np} holds) before it satisfies termination condition at Lines 16-17, it judiciously switches to iterative sequential pushes (Lines 19-28).
In each iteration of sequential pushes, \bwdp performs push operations for all nodes $u_i\in U$ and $v_i\in V$ with positive residues as Lines 4-12 in Algorithm \ref{alg:bp} except that $\epsilon_b$ at Line 4 is replaced by $0$ (Lines 20-28). The iterative process stops and returns $\{\bppr(u_i,u)|u_i\in U\}$ with $\bresidue_u$ (Line 29) when $\bresidue_{u}(u_i)\le \epsilon_b\ \forall{u_i\in U}$ or $\sum_{u_i\in U}{\bresidue_u(u_i)}\le\epsilon_b$ (Line 19). In particular, Algorithm \ref{alg:bwd} returns the residue vector $\bresidue_u$ to facilitate its combination with \fwdp, as detailed in follow-up sections. 

\subsubsection{\bf Analysis}
Theorem \ref{lem:bwd} indicates that Algorithm \ref{alg:bwd} ensures at most $\epsilon_b$ absolute error in each approximate HPP $\bppr(u_i,u)$.
\begin{theorem}[Correctness of \bwdp]\label{lem:bwd}
Given a target node $u$, \bwdp returns an approximate HPP value $\appr(u_i,u)$ for each node $u_i\in U$ such that
\begin{equation*}
\pi(u_i,u)-\appr(u_i,u)\le \epsilon_b.
\end{equation*}
\end{theorem}

Selective pushes terminate when Eq. \eqref{eq:bwd-np} holds, the cost incurred by this phase is hence $\textstyle O\left(|E|\cdot \left(\log{\tfrac{1}{\sum_{u_i\in U}{\bresidue_u(u_i)}}}\right)\right)$. In the course of sequential pushes, each iteration converts $\alpha$ portion of $\bresidue_u(u_i)\ \forall{u_i\in U}$ into the HPP. Assume that the number of iterations for sequential pushes is $t$. Note that in the worst case, the sequential pushes stops when $\sum_{u_i\in U}{\bresidue_u^{(t)}(u_i)}\le\epsilon_b$, where $\bresidue_u^{(t)}(u_i)$ signifies the residue after $t$ iterations of sequential pushes. Thus, we have
\begin{align*}
\textstyle \sum_{u_i\in U}{\bresidue_u^{(t)}(u_i)}=\sum_{u_j\in U}\left(1-\sum_{\ell=0}^{t}{\alpha(1-\alpha)^{\ell}}\right)\cdot\bresidue_u(u_j) \le \epsilon_b,
\end{align*}
which derives $\textstyle t = \log_{\tfrac{1}{1-\alpha}}{\tfrac{\sum_{u_i\in U}{\bresidue_u(u_i)}}{\epsilon_b}}-1$.
Each iteration (Lines 20-28) of sequential pushes involves a traversal of $G$; therefore the time cost is bounded by $O(|E|)$. Overall, the time complexity of \bwdp is $O(n_p+|E|\cdot t)$, which equals $\textstyle O(|E|\cdot \log{({1}/{\epsilon_b})})$.

\subsection{\fwdp}\label{sec:fwd}
\begin{algorithm}[!t]
\begin{small}
	\caption{\fwdp}
	\label{alg:fwd}
	\KwIn{Bipartite graph $G$, source node $u$, restart probability $\alpha$, parameter $\lambda$ , error threshold $\epsilon_f$, $\{\bppr(u_i,u)|u_i\in U\},\bresidue_{u}$.}
	\KwOut{$\{\fppr(u,u_i)|u_i\in U\}$.}
	\tcc{Selective pushes}
	Compute $\gamma$ by Eq. \eqref{eq:gamma}\;
	{\nonl Lines 2-14 are the same as Lines 3-15 in Algorithm \ref{alg:bwd} by replacing $\epsilon_b$ by $\frac{ws(u)}{ws(u_i)}\cdot \frac{\epsilon_f}{\lambda}$}\;
	\setcounter{AlgoLine}{14}
	{\Indp
    	\lIf{Inequality \eqref{eq:fwd-eps} holds $\forall{u_i}\in U$}{
    	    \Return $\{\fppr(u,u_i)|u_i\in U\}$ according to Eq. \eqref{eq:fppr}
    	}
    	\lIf{Inequality \eqref{eq:fwd-np} holds}{\bf{break}}
	}
	\tcc{Power iterations}
    \For{$u_i\in U$}{
        Compute $\fresidue_u(u_i)$ according to Eq. \eqref{eq:fresidue}\;
        Compute $\fppr(u,u_i)$ according to Eq. \eqref{eq:fppr}\;
    }
        Compute $t$ according to Eq. \eqref{eq:t}\;
        $\frp_{u} \gets$\piter($\UM, \VM,\fresidue_{u}(U), \alpha, t$)\;
    \lFor{$u_i\in U$}{
    $\fppr(u,u_i)\gets \fppr(u,u_i)+\frp_{u}(u_i)$
    }
    \Return $\{\fppr(u,u_i)|u_i\in U\}$\;
\end{small}
\end{algorithm}

\begin{lemma}\label{lem:asym}
For any two nodes ${u_i,u_j}\in U$, $\textstyle\frac{\pi(u,u_i)}{ws(u_i)}=\frac{\pi(u_i,u)}{ws(u)}$.
\end{lemma}

\fwdp capitalizes on the idea inspired by the following observation. Specifically, plugging Lemma \ref{lem:asym} into Lemma \ref{lem:bwd-eq} yields
\begin{small}
\begin{align}
\pi(u,u_i)&\textstyle=\frac{ws(u_i)}{ws(u)}\bppr(u_i,u)+\sum_{u_j\in U}{\pi(u_i,u_j) \bresidue_u(u_j) \frac{ws(u_i)}{ws(u)}}\nonumber\\
&\textstyle=\frac{ws(u_i)}{ws(u)}\bppr(u_i,u)+\sum_{u_j\in U}{\pi(u_j,u_i)\frac{ws(u_j)}{ws(u)} \bresidue_u(u_j)}.\label{eq:bwd2fwd}
\end{align}
\end{small}
Given an error threshold $\tfrac{\epsilon_f}{\lambda}$, if we ensure $\tfrac{ws(u_j)}{ws(u)} \cdot \bresidue_u(u_j)\le \tfrac{\epsilon_f}{\lambda}$, Eq. \eqref{eq:bwd2fwd} becomes $\pi(u,u_i)\le\tfrac{ws(u_i)}{ws(u)}\cdot\bppr(u_i,u)+\tfrac{\epsilon_f\cdot\sum_{u_j\in U}{\pi(u_j,u_i)}}{\lambda}$, which implies that $\tfrac{ws(u_i)}{ws(u)}\cdot\bppr(u_i,u)$ can be regarded as an underestimate of $\pi(u,u_i)$. In this regard, instead of employing \piter to compute the HPP values w.r.t. a source node $u$ from scratch, we can transform the residues and the approximate HPP values returned by \bwdp to obtain a rough approximation, and further refine it in \fwdp using a few selective pushes and power iterations. Along this line, we summarize the pseudo-code of \fwdp in Algorithm \ref{alg:fwd} and explain its details in the sequel.

\subsubsection{\bf Details}
The input parameters of \fwdp are identical to those of \bwdp, except that it accepts absolute error threshold $\epsilon_f$ instead of $\epsilon_b$, an additional parameter $\lambda$, as well as the initial approximate HPP values $\{\bppr(u_i,u)|u_i\in U\}$ and residue vector $\bresidue_{u}$ returned by \bwdp. In analogy to \bwdp, \fwdp consists of two phases: selective pushes (Lines 1-16) and power iterations (Lines 17-22), where power iterations are used to mitigate the efficiency issue of selective pushes when excessive push operations occur, as done in Section \ref{sec:bwd}. First, \fwdp iteratively executes selective pushes as Lines 3-15 in Algorithm \ref{alg:bwd}. Distinct from \bwdp, the selective threshold $\epsilon_b$ is replaced by $\frac{ws(u)}{ws(u_i)}\cdot \frac{\epsilon_f}{\lambda}$. In addition, the termination conditions of selective pushes in Algorithm \ref{alg:fwd} are changed as follows: (i) for each node $u_i\in U$,
\begin{equation}\label{eq:fwd-eps}
\textstyle \bresidue_{u}(u_i)\le \frac{ws(u)}{ws(u_i)}\cdot \frac{\epsilon_f}{\lambda}
\end{equation}
holds, or (ii) the actual cost incurred by selective pushes $n_p$ exceeds a pre-defined computation budget, {\it viz.},
\begin{small}
\begin{equation}\label{eq:fwd-np}
\textstyle n_p\ge 2|E|\cdot \left(\log_{\frac{1}{1-\alpha}}{\frac{\gamma}{\sum_{u_i\in U}{\frac{ws(u_i)}{ws(u)}\cdot\bresidue_u(u_i)}}}\right),
\end{equation}
\end{small}
where $\gamma$ is computed at Line 1 based on Equation \ref{eq:gamma}.
\begin{equation}\label{eq:gamma}
\textstyle \gamma\gets\sum_{u_i\in U}{\frac{ws(u_i)}{ws(u)}\cdot\bresidue_u(u_i)}.
\end{equation}
If the first condition holds, Algorithm \ref{alg:fwd} returns 
\begin{equation}\label{eq:fppr}
\textstyle\fppr(u,u_i)= \frac{ws(u_i)}{ws(u)}\cdot \bppr(u_i,u)
\end{equation}
as an estimation of $\ppr(u,u_i)$ for each node $u_i\in U$ (Line 15). On the other hand, if Inequality \eqref{eq:fwd-np} (\ie the second condition) is satisfied (Line 16), \fwdp proceeds to refine the result with a few power iterations. More specifically, we first transform the approximate HPP values $\bppr(u_i,u)$ and residues $\bresidue_u(u_i)$ into the approximate HPP values $\fppr(u,u_i)$ and residues $\fresidue_u(u_i)$ according to Eq. \eqref{eq:fppr} and Eq. \eqref{eq:fresidue}, respectively (Lines 17-19).
\begin{equation}\label{eq:fresidue}
\fresidue_u(u_i)= \textstyle\frac{ws(u_i)}{ws(u)}\cdot\bresidue_u(u_i)
\end{equation}
Subsequently, Algorithm \ref{alg:fwd} computes $\frp_{u}(u_i)$ as an additional estimation part of $\ppr(u,u_i)$ by invoking \piter with input parameters including graph $G$, a length-$|U|$ residue vector $\fresidue_u(U)$, restart probability $\alpha$, and the number of iterations $t$ defined in Eq. \eqref{eq:t} (Line 21).
\begin{small}
\begin{equation}\label{eq:t}
\textstyle t=\log_{\frac{1}{1-\alpha}}{\frac{\sum_{u_i\in U}{\fresidue_u(u_i)}}{\epsilon_f}}-1
\end{equation}
\end{small}
Eventually, \fwdp gives each $\fppr(u,u_i)$ a final touch by adding $\frp_{u}(u_i)$ to it and returns it as an estimation of $\ppr(u,u_i)$ (Lines 22-23).

\begin{algorithm}[!t]
\begin{small}
	\caption{\algo}
	\label{alg:main}
	\KwIn{Bipartite graph $G$, query node $u$, restart probability $\alpha$, parameter $\lambda$, and error thresholds $\epsilon,\epsilon_b$.}
	\KwOut{$\{\beta^{\prime}(u,u_i)|u_i\in U\}$.}
    $\{\bppr(u_i,u)|u_i\in U\},\bresidue_{u} \gets\mathtt{\bwdt}(G,\alpha,u,\epsilon_b)$\;
    $\epsilon_f \gets \epsilon-\epsilon_b$\;
    $\{\fppr(u,u_i)|u_i\in U\} \gets\mathtt{\fwdt}(G,\alpha,u,\epsilon_f,\lambda,\{\bppr(u_i,u)|u_i\in U\},\bresidue_u)$\;
    \lFor{$u_i\in U$}{$\beta^{\prime}(u,u_i)\gets \fppr(u,u_i)+\bppr(u_i,u)$}
	\Return $\{\beta^{\prime}(u,u_i)|u_i\in U\}$\;
\end{small}
\end{algorithm}

\subsubsection{\bf Analysis}
Theorem \ref{lem:fwd} establishes the accuracy guarantees of \fwdp.
\begin{theorem}[Correctness of \fwdp]\label{lem:fwd}
Algorithm \ref{alg:fwd} returns an approximate HPP $\appr(u,u_i)$ for each node $u_i\in U$ such that
$\pi(u,u_i)-\appr(u,u_i)\le \epsilon_f$, when the input parameter $\lambda$ satisfies
\begin{equation}\label{eq:lambda-ineq}
\textstyle \lambda \ge \max_{u_i\in U}{\sum_{u_j\in U}{\pi(u_j,u_i)}}.
\end{equation}
\end{theorem}

Next, we analyse the time complexity of \fwdp. First, according to Line 16 in Algorithm \ref{alg:fwd}, the selective pushes in \fwdp take $O\textstyle\left(|E|\cdot \left(\log{\frac{\gamma}{\sum_{u_i\in U}{\fresidue_u(u_i)}}}\right)\right)$ time. Moreover, each iteration in \piter consumes $O(|E|)$ time and \fwdp performs $t$ (see Eq. \eqref{eq:t}) power iterations in total (Line 21 in Algorithm \ref{alg:fwd}). Consequently, the overall time complexity of \fwdp is
$O\left(|E|\cdot \left(\log{\frac{\gamma}{\epsilon_f}}\right)\right)$.

\subsection{Complete Algorithm and Analysis}
Algorithm \ref{alg:main} summarizes the pseudo-code of \algo, which takes as input a bipartite graph $G$, query node $u$, restart probability $\alpha$, two error thresholds $\epsilon,\epsilon_b$ ($\epsilon_b<\epsilon$), as well as a parameter $\lambda$ (see Eq. \eqref{eq:lambda-ineq}). In particular, $\lambda$ can be efficiently estimated in the {\em preprocessing} step based on Eq. \eqref{eq:lambda} and is guaranteed to be a tight upper bound of $\max_{u_i\in U}{\sum_{u_j\in U}{\pi(u_j,u_i)}}$ based on Lemma \ref{lem:lambda}.
\begin{lemma}\label{lem:lambda}
Suppose that $\boldsymbol{\rho}$ be the result of \piter when the input parameters $\mathbf{e}=\mathbf{1}$ and $t=\tau$. Then, we have $\lambda\ge \max_{u_i\in U}{\sum_{u_j\in U}{\pi(u_j,u_i)}}$ holds, where 
\begin{small}
\begin{equation}\label{eq:lambda}
\textstyle \lambda=\min\left\{\max_{u_i\in U}{\boldsymbol{\rho}(u_i)}+|U|\cdot (1-\alpha)^{\tau+1}, \frac{\max_{u_i\in U}{ws(u_i)}}{\min_{u_j\in U}{ws(u_j)}}\right\}.
\end{equation}
\end{small}
\end{lemma}

In online phase, Algorithm \ref{alg:main} first invokes \bwdp with absolute error threshold $\epsilon_b$ to compute $\{\bppr(u_i,u)|u_i\in U\}$ and $\bresidue_{u}$ (Line 1). Then, these intermediate results together with $\epsilon_f=\epsilon-\epsilon_b$ are passed over to \fwdp. After the invocation of \fwdp, we obtain $\{\fppr(u,u_i)|u_i\in U\}$. Finally, the approximate BHPP value for each node $u_i$ in $U$ can be calculated by $\beta^{\prime}(u,u_i)=\fppr(u,u_i)+\bppr(u_i,u)$, which is ensured to be $\epsilon$-approximate, as shown in  Theorem \ref{lem:main}.

\begin{theorem}[Correctness of \algo]\label{lem:main}
Algorithm \ref{alg:main} returns $\beta^{\prime}(u,u_i)$ for every $u_i\in U$ that satisfies Eq. \eqref{eq:eps-BHPP}.
\end{theorem}

\begin{figure}[!t]
\centering
\begin{small}
\begin{tikzpicture}
\begin{axis}[
    height=\textwidth/5.5,
    width=\textwidth/3.6,
    ylabel=$\gamma$,
    xlabel=$\epsilon_b$,
    xmin=0, xmax=0.1,
    ymin=0, ymax=1.0,
    xtick={0,0.02,0.04,0.06,0.08,0.1},
    xticklabels={0,0.02,0.04,0.06,0.08,0.1},
    every axis x label/.style={at={(current axis.right of origin)},above=0mm,left=-2mm,anchor=north west},
    legend style={at={(1.5,0.85)},anchor=north,draw=none,font=\small} 
    ]
    \addplot[
        line width=0.7pt,
        scatter/classes={Avito={mark=o}, DBLP={mark=square}, KDDCup={mark=triangle}},
        scatter, 
        scatter src=explicit symbolic,
    ] table [meta=class] {
    x y class
0.5	0.816536 Avito
0.2	0.705626 Avito
0.1	0.614815 Avito
0.05	0.520884 Avito
0.02	0.43058 Avito
0.01	0.373613 Avito
0.005	0.31299 Avito
0.002	0.228509 Avito
0.001	0.174406 Avito
0.0005	0.128977 Avito
0.0002	0.0866376 Avito
0.0001	0.0648221 Avito
0.00005	0.0487735 Avito
0.00002	0.033657 Avito
0.00001	0.0254703 Avito
0.000005	0.0191131 Avito
0.000002	0.0125269 Avito
0.000001	0.00873584 Avito
0.0000005	0.00595517 Avito
0.0000002	0.00343788 Avito
0.0000001	0.00219468 Avito

0.5	0.85 DBLP
0.2	0.848172 DBLP
0.1	0.846692 DBLP
0.05	0.844154 DBLP
0.02	0.840402 DBLP
0.01	0.832079 DBLP
0.005	0.817314 DBLP
0.002	0.791884 DBLP
0.001	0.757893 DBLP
0.0005	0.711808 DBLP
0.0002	0.612483 DBLP
0.0001	0.494539 DBLP
0.00005	0.372045 DBLP
0.00002	0.221549 DBLP
0.00001	0.133969 DBLP
0.000005	0.0713802 DBLP
0.000002	0.0324092 DBLP
0.000001	0.0155047 DBLP
0.0000005	0.00845155 DBLP
0.0000002	0.00363672 DBLP
0.0000001	0.00176441 DBLP
		

0.5	0.625829 KDDCup
0.2	0.438092 KDDCup
0.1	0.344758 KDDCup
0.05	0.292877 KDDCup
0.02	0.22566 KDDCup
0.01	0.191414 KDDCup
0.005	0.163711 KDDCup
0.002	0.13095 KDDCup
0.001	0.10991 KDDCup
0.0005	0.0911064 KDDCup
0.0002	0.0670996 KDDCup
0.0001	0.0492659 KDDCup
0.00005	0.0356114 KDDCup
0.00002	0.0236315 KDDCup
0.00001	0.0176909 KDDCup
0.000005	0.0134709 KDDCup
0.000002	0.00936593 KDDCup
0.000001	0.0071669 KDDCup
0.0000005	0.00528755 KDDCup
0.0000002	0.0034904 KDDCup
0.0000001	0.00250058 KDDCup
};
\legend{Avito,DBLP,KDDCup}
\end{axis}
\end{tikzpicture}
\end{small}
\vspace{-2mm}
\caption{$\gamma$ vs. $\epsilon_b$.} \label{fig:gamma}
\vspace{0mm}
\end{figure}
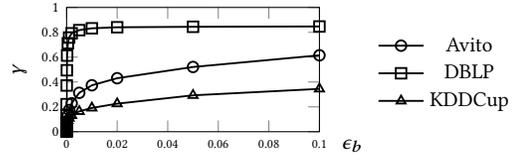

\subsubsection{\bf Choosing $\boldsymbol{\epsilon_b}$ and Time Complexity.}\label{sec:epsilon-b}In the following, we discuss how to pick $\epsilon_b$ reasonably such that \algo runs in optimal time.
Since $\epsilon_f+\epsilon_b=\epsilon$, the total computational time of \algo can be formulated as the following function:
\begin{small}
\begin{align}
\textstyle f(\epsilon_b)&\textstyle=O\left(|E|\cdot \left(\log{\frac{1}{\epsilon_b}}+\log{\frac{\gamma}{\epsilon-\epsilon_b}}\right)\right)\textstyle=O\left(|E|\cdot \left(\log{\frac{\gamma}{\epsilon_b\cdot (\epsilon-\epsilon_b) }}\right)\right)\nonumber.
\end{align}
\end{small}
Thus, \algo's time complexity is bounded by $\textstyle O\left(|E|\cdot \log{\frac{1}{\epsilon }}\right)$ since $\gamma\le 1$. According to $f(\epsilon_b)$, as $\epsilon_b$ decreases and \bwdp progresses, $\gamma$ monotonically decreases and the cost of \fwdp reduces. To strike a balance between the costs incurred by \bwdp and \fwdp, we need to choose an appropriate $\epsilon_b$. For this purpose, we first quantify the relationship between $\gamma$ and $\epsilon_b$ based on the following empirical finding.
Figure \ref{fig:gamma} plots the $\gamma$ values when varying $\epsilon_b$ from $10^{-6}$ to $0.1$ on three real datasets used in our experiments in Section \ref{sec:exp}. It can be observed that $\gamma \propto \epsilon_b^{\mu}$ ($0<\mu\le 1$), where $\mu$ is roughly $\textstyle{\sqrt{|U|\cdot|V|}}/{|E|}$. Consequently, the computational time of \algo can be minimized by 
\begin{small}
\begin{equation}\label{eq:epsilon-b}
\textstyle \epsilon_b=\epsilon\cdot\frac{1-\mu}{2-\mu}=\epsilon\cdot \frac{|E|-{\sqrt{|U|\cdot|V|}}}{2|E|-{\sqrt{|U|\cdot|V|}}}.
\end{equation}
\end{small}

\section{Experiments}\label{sec:exp}
This section experimentally evaluates the effectiveness of BHPP and other eight similarity measures on query rewriting and item recommendation on 7 real datasets, and compares \algo against two baseline solutions for $\epsilon$-approximate BHPP queries in terms of efficiency. For a fair comparison, all algorithms are implemented in C++ and compiled by g++ 7.5 with $\mathtt{-O3}$ optimization, and all experiments are conducted on a Linux machine with an Intel Xeon(R) E7-8880 v4@2.20GHz CPU and 1TB RAM.

\begin{table}[t]
\centering
\renewcommand{\arraystretch}{1.0}
\begin{footnotesize}
\caption{Statistics of click graphs.}\label{tbl:exp-click-data}\vspace{-2mm}
\resizebox{\columnwidth}{!}{%
\begin{tabular}{|l|r|r|r|r|c|c|}
	\hline
	{\bf Name} & \multicolumn{1}{c|}{$|U|$} & \multicolumn{1}{c|}{$|V|$} & \multicolumn{1}{c|}{$|E|$} & \multicolumn{1}{c|}{\#clicks}& \multicolumn{1}{c|}{\#impressions} \\
	\hline
	{\bf\em Avito} \cite{avito}  & 27,736 & 16,589 & 67,028 & 278,960 & 18,121,561 \\ \hline
	{\bf\em KDDCup} \cite{kddcup} & 255,170 & 1,848,114 & 2,766,393 & 8,217,633 & 121,232,353 \\ \hline
	{\bf\em AOL} \cite{aol} & 4,811,647 & 1,632,788 & 10,741,953 & 19,442,625 & 69,745,428,949 \\ \hline
\end{tabular}%
}
\end{footnotesize}
\vspace{0mm}
\end{table}

\begin{table}[t]
\centering
\renewcommand{\arraystretch}{1.0}
\begin{footnotesize}
\caption{Statistics of user-item graphs.}\label{tbl:exp-data}\vspace{-2mm}
\begin{tabular}{|l|r|r|r|c|}
	\hline
	{\bf Name} & \multicolumn{1}{c|}{$|V|$} & \multicolumn{1}{c|}{$|U|$} & \multicolumn{1}{c|}{$|E|$}& \multicolumn{1}{c|}{\bf weight} \\
	\hline
	{\bf\em DBLP} \cite{BiNE18} & 6,001 & 1,308 & 29,256 & \#papers \\
	\hline	
	{\bf\em MovieLens} \cite{m1-1m} & 6,040  & 3,706 & 1,000,209 & ratings\\
	\hline
 	{\bf\em Last.fm} \cite{lastfm} & 359,349 & 160,168 & 17,559,530 & \#plays\\
 	\hline	
 	{\bf\em Amazon-Games} \cite{amazon-games}  & 826,767 & 50,210 & 1,324,753  & ratings \\
	\hline
\end{tabular}%
\end{footnotesize}
\vspace{0mm}
\end{table}

\subsection{Datasets}
We experiment with 7 real bipartite graphs that are used in previous work \cite{deng2009generalized,harper2015movielens,epasto2014reduce,pass2006picture,he2016ups,mcauley2015image,Celma:Springer2010}, including 3 click graphs and 4 user-item graphs, which will be used in query rewriting and item recommendation in Sections \ref{sec:qr} and \ref{sec:ir}, respectively.

\subsubsection{\bf Click graphs.}
Table \ref{tbl:exp-click-data} lists the statistics of the 3 click graphs: {\em Avito}, {\em KDDCup}, and {\em AOL}, each of which is a bipartite graph containing a set of queries in $U$ and a set of ads/URLs in $V$. Each edge in them connects a query with
an ad/URL if and only if the ad/URL was clicked at least once by a user who issued the query. Additionally, each edge
is associated with two numbers, \ie \#clicks and \#impressions, which signify the number of clicks that the ad/URL received and the displayed times of the ad/URL, respectively. The weight of each edge is the ratio of its \#clicks to its \#impressions.

\subsubsection{\bf User-item graphs.}
The statistics of 4 user-item graphs including {\em DBLP}, {\em MovieLens}, {\em Last.fm}, and {\em Amazon-Games} are reported in Table \ref{tbl:exp-data}, where $U$ (resp. $V$) contains items (resp. users) and $E$ is the set of interactions between users and items. The weight of an edge in the publication network {\em DBLP} indicates the number of papers published on a venue by an author. 
{\em MovieLens} is a well-known dataset for recommender systems, in which each edge weight denotes a user's rating on a movie. In {\em Last.fm}, each edge is associated with a weight representing the play count of a song by a user. 
{\em Amazon-Games} is extracted from the Amazon review data, in which each edge weight is a rating of a game assigned by a user. To ensure the dataset quality, we apply the 10-core setting \cite{he2016vbpr} on the graphs, \ie removing users and items with less than ten edges.

\begin{table*}[ht]
\centering
\renewcommand{\arraystretch}{1.0}
\caption{Item recommendation performance ($k=10$).}\vspace{-2mm}
\begin{footnotesize}
\begin{tabular}{|c|c c|c c|c c| c c|}
\hline
\multirow{2}{*}{\bf Similarity} & \multicolumn{2}{c|}{\bf {\em DBLP}} & \multicolumn{2}{c|}{\bf {\em Movielens}} & \multicolumn{2}{c|}{\bf {\em Last.fm}} &
\multicolumn{2}{c|}{\bf {\em Amazon-Games}} \\ \cline{2-9}
& $precision@k$ & $recall@k$ & $precision@k$ & $recall@k$ & $precision@k$ & $recall@k$ &  $precision@k$ & $recall@k$ \\ 
\hline
BHPP & {\bf 0.167} & {\bf 0.164} & {\bf 0.405} & {\bf 0.289} & {\bf 0.313} & {\bf 0.231} & {\bf 0.248}  & {\bf 0.187} \\ 
\hline
HPP  & 0.14 & 0.138 & 0.224 & 0.161 & 0.305 & 0.223 & 0.194  & 0.15 \\ 
Pearson  & 0.037 & 0.037 & 0.106 & 0.074 & 0.126 & 0.095 & 0.056  & 0.044 \\ 
Jaccard  & 0.158 & 0.157 & 0.272 & 0.194 & 0.287 & 0.213 & 0.08  & 0.062 \\ 
SimRank  & 0.151 & 0.15 & 0.245 & 0.177 & 0.239 & 0.169 & 0.127  & 0.084 \\ 
CoSimRank  & 0.115 & 0.113 & 0.186 & 0.137 & 0.304 & 0.216 &  0.156  & 0.121 \\ 
PPR  & 0.149 & 0.146 & 0.342 & 0.245 & 0.28 & 0.206 &  0.188  & 0.143 \\ 
SimRank++  & 0.127 & 0.126 & 0.243 & 0.176 & 0.241 & 0.171  & 0.171 &  0.118 \\ 
P-SimRank  & 0.127 & 0.127 & 0.221 & 0.164 & 0.226 & 0.159  & 0.14 & 0.088 \\ 
\hline
\end{tabular}
\end{footnotesize}
\label{tab:ir-acc-10}
\vspace{-1mm}
\end{table*}

\subsection{Query Rewriting}\label{sec:qr}

\header
{\bf Settings.}
Given a click graph and a query $q_i$ in the graph, query rewriting aims to transform $q_i$ into equivalent queries based on their similarities. First, following prior work \cite{antonellis2008simrank++}, we calculate the {\em desirability} of query $q_j$ w.r.t. query $q_i$ on the original click graph as $\textstyle  des(q_i,q_j)=\sum_{q_k\in {N}(q_i)\cap {N}(q_j)}{\frac{w(q_j,q_k)}{{d}(q_j)}}$.
To evaluate the effectiveness of BHPP and competing similarity measures on query rewriting, we remove $20\%$ edges from the original click graph and on the remaining graph compute similarity scores between queries based on the definition of each similarity measure. As in previous work \cite{deng2009generalized,antonellis2008simrank++}, we test whether the top-$k$ (typically, $k=10$ and $5$) ordering of queries yielded by the similarity scores is consistent with the one derived from the desirability scores, using the classic evaluation metric {\em Normalized Discounted Cumulative Gain} (NDCG) \cite{jarvelin2017ir}. The competing similarity measures include Pearson's correlation coefficient \cite{antonellis2008simrank++}, Jaccard's coefficient \cite{jaccard1912distribution}, SimRank \cite{jeh2002simrank}, naive PPR \cite{haveliwala2002topic,jeh2003scaling}, CoSimRank \cite{rothe2014cosimrank}, SimRank++ \cite{antonellis2008simrank++}, P-SimRank \cite{dey2020p} and HPP \cite{deng2009generalized,epasto2014reduce}. For each dataset, we carry out query rewriting tasks for $100$ randomly selected queries, and report the average performance, \ie NDCG scores.

\header
{\bf Results.} Figure \ref{fig:qr-10} shows the NDCG results of BHPP and other eight similarity measures on {\em Avito}, {\em KDDCup}, and {\em AOL} datasets when $k=10$. We can see that BHPP consistently outperforms other similarity measures on three datasets. In particular, on {\em Avito}, BHPP achieves a remarkable improvement of at least $2\%$ over state-of-the-art results. On {\em KDDCup}, BHPP is superior to competing similarity measures with a considerable gain of up to $0.4\%$. From Figure \ref{fig:qr-10}(c), we can see that BHPP is slightly better than the best competitor HPP on {\em AOL}. 
Similar observations can be made from the experimental outcomes when $k=5$.
For the interest of space, we refer interested readers to Appendix for the detailed results.

\begin{figure}[!t]
\centering
\begin{small}
\begin{tikzpicture}
    \begin{customlegend}[legend columns=5,
        legend entries={BHPP,HPP,Pearson,Jaccard,SimRank},
        area legend,
        legend style={at={(0.45,1.15)},anchor=north,draw=none,font=\scriptsize,column sep=0.15cm}]
        \addlegendimage{,fill=black}
        \addlegendimage{,pattern=crosshatch dots}
        \addlegendimage{,pattern=grid}
        \addlegendimage{,pattern=sixpointed stars}
        \addlegendimage{,pattern=vertical lines}
    \end{customlegend}
\end{tikzpicture}
\\[-\lineskip]
\vspace{-1mm}
\begin{tikzpicture}
    \begin{customlegend}[legend columns=4,
        legend entries={PPR,CoSimRank,SimRank++,P-SimRank},
        area legend,
        legend style={at={(0.45,1.15)},anchor=north,draw=none,font=\scriptsize,column sep=0.15cm}]
        \addlegendimage{,pattern=dots}
        \addlegendimage{,pattern=horizontal lines}
        \addlegendimage{,pattern=fivepointed stars}
        \addlegendimage{,pattern=north east lines}
    \end{customlegend}
\end{tikzpicture}
\\[-\lineskip]
\vspace{-3mm}
\subfloat[{\em Avito}]{
\begin{tikzpicture}[scale=1]
\begin{axis}[
    height=\columnwidth/2.7,
    width=\columnwidth/2.3,
    ybar=0.2pt,
    bar width=0.22cm,
    enlarge x limits=true,
    ylabel={\em NDCG}$@k$,
    xticklabel=\empty,
    ymin=0.75,
    ymax=0.85,
    ytick={0.75,0.77,0.79,0.81,0.83,0.85},
    every axis y label/.style={font=\scriptsize,at={(current axis.north west)},right=4.5mm,above=0mm},
    legend style={at={(0.02,0.98)},anchor=north west,cells={anchor=west},font=\tiny}
    ]

\addplot [,fill=black] coordinates {(1,0.841406) }; 
\addplot [,pattern=crosshatch dots] coordinates {(1,0.778)}; 
\addplot [,pattern=grid] coordinates {(1,0.767018)}; 
\addplot [,pattern=sixpointed stars] coordinates {(1,0.773591)}; 
\addplot [,pattern=vertical lines] coordinates {(1,0.787330)}; 
\addplot [,pattern=horizontal lines] coordinates {(1,0.820839)}; 
\addplot [,pattern=dots] coordinates {(1,0.817299)}; 
\addplot [,pattern=fivepointed stars] coordinates {(1,0.787608)}; 
\addplot [,pattern=north east lines] coordinates {(1,0.797164)}; 
\end{axis}
\end{tikzpicture}\vspace{-2mm}\hspace{0mm}%
}%
\subfloat[{\em KDDCup}]{
\begin{tikzpicture}[scale=1]
\begin{axis}[
    height=\columnwidth/2.7,
    width=\columnwidth/2.3,
    ybar=0.2pt,
    bar width=0.22cm,
    enlarge x limits=true,
    ylabel={\em NDCG}$@k$,
    xticklabel=\empty,
    ymin=0.986,
    ymax=0.998,
    ytick={0.986,0.990,0.994,0.998},
    yticklabels={0.986,0.990,0.994,0.998},
    every axis y label/.style={font=\scriptsize,at={(current axis.north west)},right=4.5mm,above=0mm},
    legend style={at={(0.02,0.98)},anchor=north west,cells={anchor=west},font=\tiny}
    ]

\addplot [,fill=black] coordinates {(1,0.99625) }; 
\addplot [,pattern=crosshatch dots] coordinates {(1,0.991)}; 
\addplot [,pattern=grid] coordinates {(1,0.989042)}; 
\addplot [,pattern=sixpointed stars] coordinates {(1,0.991126)}; 
\addplot [,pattern=vertical lines] coordinates {(1,0.99295)}; 
\addplot [,pattern=horizontal lines] coordinates {(1,0.989763)}; 
\addplot [,pattern=dots] coordinates {(1,0.993007)}; 
\addplot [,pattern=fivepointed stars] coordinates {(1,0.99295)}; 
\addplot [,pattern=north east lines] coordinates {(1,0.99431)}; 
\end{axis}
\end{tikzpicture}\vspace{-2mm}\hspace{0mm}%
}%
\subfloat[{\em AOL}]{
\begin{tikzpicture}[scale=1]
\begin{axis}[
    height=\columnwidth/2.7,
    width=\columnwidth/2.3,
    ybar=0.2pt,
    bar width=0.22cm,
    enlarge x limits=true,
    ylabel={\em NDCG}$@k$,
    xticklabel=\empty,
    ymin=0.99,
    ymax=0.996,
    ytick={0.99,0.992,0.994,0.996},
    yticklabels={0.99,0.992,0.994,0.996},
    every axis y label/.style={font=\scriptsize,at={(current axis.north west)},right=4.5mm,above=0mm},
    legend style={at={(0.02,0.98)},anchor=north west,cells={anchor=west},font=\tiny}
    ]

\addplot [,fill=black] coordinates {(1,0.995233) }; 
\addplot [,pattern=crosshatch dots] coordinates {(1,0.995)}; 
\addplot [,pattern=grid] coordinates {(1,0.991647)}; 
\addplot [,pattern=sixpointed stars] coordinates {(1,0.99118)}; 
\addplot [,pattern=vertical lines] coordinates {(1,0.993961)}; 
\addplot [,pattern=horizontal lines] coordinates {(1,0.994006)}; 
\addplot [,pattern=dots] coordinates {(1,0.9931)}; 
\addplot [,pattern=fivepointed stars] coordinates {(1,0.993961)}; 
\addplot [,pattern=north east lines] coordinates {(1,0.994060)}; 
\end{axis}
\end{tikzpicture}\vspace{-2mm}\hspace{0mm}%
}%
\vspace{-3mm}
\end{small}
\caption{Query rewriting performance ($k=10$).} \label{fig:qr-10}
\vspace{0mm}
\end{figure}
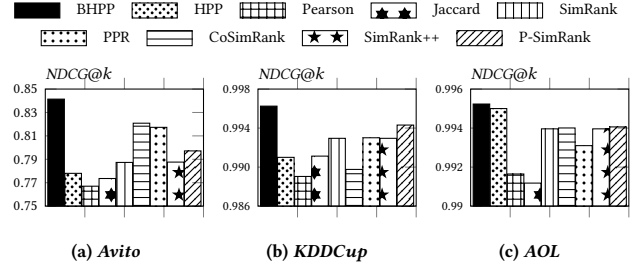

\begin{figure}[!t]
\centering
\begin{small}
\begin{tikzpicture}
    \begin{customlegend}[legend columns=3,
        legend entries={\algo,\mcsp,\pisp},
        legend style={at={(0.45,1.05)},anchor=north,draw=none,font=\scriptsize,column sep=0.1cm}]
    \addlegendimage{line width=0.25mm,mark=triangle}
    \addlegendimage{line width=0.25mm,mark=square}
    \addlegendimage{line width=0.25mm,mark=o}
    \end{customlegend}
\end{tikzpicture}
\\[-\lineskip]
\vspace{-3mm}
\subfloat[{\em DBLP}]{
\begin{tikzpicture}[scale=1]
    \begin{axis}[
        height=\columnwidth/2.7,
        width=\columnwidth/1.9,
        ylabel={\em query time (ms)},
        xlabel=$\epsilon$,
        xmin=0.5, xmax=6.5,
        ymin=1, ymax=100000,
        xtick={1,2,3,4,5,6},
        xticklabel style = {font=\scriptsize},
        yticklabel style = {font=\scriptsize},
        xticklabels={1e-2,1e-3,1e-4,1e-5,1e-6,1e-7},
        scaled y ticks = false,
        ymode=log,
        log basis y={10},
        ytick = {1,10,100,1000,10000,100000},
        every axis y label/.style={at={(current axis.north west)},right=9mm,above=0mm},
        every axis x label/.style={at={(current axis.right of origin)},above=0mm,anchor=north west},
    ]
    \addplot[line width=0.25mm,mark=triangle]  
        plot coordinates {
(1,	2.02152)
(2,	24.4367)
(3,	63.3597)
(4,	96.695)
(5,	126.364)
(6,	160.264)
    };

    \addplot[line width=0.25mm,mark=square]  
        plot coordinates {
(1,	6046.931)
(2,	34981.5)
(3,	0)
(4,	0)
(5,	0)
(6,	0)
    };
    
    \addplot[line width=0.25mm,mark=o]  
        plot coordinates {
(1,	149.934)
(2,	210.478)
(3,	287.894)
(4,	373.275)
(5,	479.015)
(6, 569.533)
    };
    \end{axis}
\end{tikzpicture}\hspace{4mm}\vspace{-1mm}%
}%
\subfloat[{\em MovieLens}]{
\begin{tikzpicture}[scale=1]
    \begin{axis}[
        height=\columnwidth/2.7,
        width=\columnwidth/1.9,
        ylabel={\em query time (ms)},
        xlabel=$\epsilon$,
        xmin=0.5, xmax=6.5,
        ymin=10, ymax=1000000,
        xtick={1,2,3,4,5,6},
        xticklabel style = {font=\scriptsize},
        yticklabel style = {font=\scriptsize},
        xticklabels={1e-2,1e-3,1e-4,1e-5,1e-6,1e-7},
        scaled y ticks = false,
        ymode=log,
        log basis y={10},
        ytick = {1,10,100,1000,10000,100000,1000000},
        every axis y label/.style={at={(current axis.north west)},right=9mm,above=0mm},
        every axis x label/.style={at={(current axis.right of origin)},above=3mm,anchor=north west},
    ]
    \addplot[line width=0.25mm,mark=triangle]  
        plot coordinates {
(1,	22.6553	)
(2,	955.43	)
(3,	2433.06	)
(4,	3808.06	)
(5,	5238.9	)
(6, 6716.91)
    };

    \addplot[line width=0.25mm,mark=square]  
        plot coordinates {
(1,	8096.464	)
(2,	110327	)
(3,	0	)
(4,	0	)
(5,	0	)
(6,	0	)
    };
    
    \addplot[line width=0.25mm,mark=o]  
        plot coordinates {
(1,	7473.73	)
(2,	10873.3	)
(3,	14744	)
(4,	19496.6	)
(5,	23875.8	)
(6, 28814.8)
    };
    
    \end{axis}
\end{tikzpicture}\hspace{4mm}\vspace{-1mm}%
}%

\subfloat[{\em KDDCup}]{
\begin{tikzpicture}[scale=1]
    \begin{axis}[
        height=\columnwidth/2.7,
        width=\columnwidth/1.9,
        ylabel={\em query time (ms)},
        xlabel=$\epsilon$,
        xmin=0.5, xmax=6.5,
        ymin=100, ymax=1000000,
        xtick={1,2,3,4,5,6},
        xticklabel style = {font=\scriptsize},
        yticklabel style = {font=\scriptsize},
        xticklabels={1e-2,1e-3,1e-4,1e-5,1e-6,1e-7},
        scaled y ticks = false,
        ymode=log,
        log basis y={10},
        ytick = {100,1000,10000,100000,1000000},
        every axis y label/.style={at={(current axis.north west)},right=9mm,above=0mm},
        every axis x label/.style={at={(current axis.right of origin)},above=8mm,anchor=north west},
    ]
    \addplot[line width=0.25mm,mark=triangle]  
        plot coordinates {
(1,	326.762)
(2,	1615.09	)
(3,	5800.46	)
(4,	15098.5	)
(5,	30064.3	)
(6,	59922.2	)
    };

    \addplot[line width=0.25mm,mark=square]  
        plot coordinates {
(1,	9017.277	)
(2,	50528.1	)
(3, 551000	)
(4,	0	)
(5,	0	)
(6, 0   )
    };
    
    \addplot[line width=0.25mm,mark=o]  
        plot coordinates {
(1,	53383.1	)
(2,	81201.4	)
(3,	106207	)
(4,	134261	)
(5,	163289	)
(6,	181957	)
    };
    
    \end{axis}
\end{tikzpicture}\hspace{4mm}\vspace{-1mm}%
}%
\subfloat[{\em Amazon-Games}]{
\begin{tikzpicture}[scale=1]
    \begin{axis}[
        height=\columnwidth/2.7,
        width=\columnwidth/1.9,
        ylabel={\em query time (ms)},
        xlabel=$\epsilon$,
        xmin=0.5, xmax=6.5,
        ymin=10, ymax=1000000,
        xtick={1,2,3,4,5,6},
        xticklabel style = {font=\scriptsize},
        yticklabel style = {font=\scriptsize},
        xticklabels={1e-2,1e-3,1e-4,1e-5,1e-6,1e-7},
        scaled y ticks = false,
        ymode=log,
        log basis y={10},
        ytick = {10,100,1000,10000,100000,1000000},
        every axis y label/.style={at={(current axis.north west)},right=9mm,above=0mm},
        every axis x label/.style={at={(current axis.right of origin)},above=3mm,anchor=north west},
    ]
    \addplot[line width=0.25mm,mark=triangle]  
        plot coordinates {
(1,	82.7281	)
(2,	545.258	)
(3,	3551.2	)
(4,	10938.6	)
(5,	20520.6	)
(6,	30387.6	)
    };

    \addplot[line width=0.25mm,mark=square]  
        plot coordinates {
(1,	12164.14	)
(2,	179867	)
(3,	0	)
(4,	0	)
(5,	0	)
(6,	0	)
    };
    
    \addplot[line width=0.25mm,mark=o]  
        plot coordinates {
(1,	34971.4	)
(2,	51371.3	)
(3,	65662	)
(4,	81759.1	)
(5,	99005.2	)
(6,	125869	)
    };
    
    \end{axis}
\end{tikzpicture}\hspace{4mm}\vspace{-1mm}%
}%

\subfloat[{\em Last.fm}]{
\begin{tikzpicture}[scale=1]
    \begin{axis}[
        height=\columnwidth/2.7,
        width=\columnwidth/1.9,
        ylabel={\em query time (ms)},
        xlabel=$\epsilon$,
        xmin=0.5, xmax=6.5,
        ymin=100, ymax=10000000,
        xtick={1,2,3,4,5,6},
        xticklabel style = {font=\scriptsize},
        yticklabel style = {font=\scriptsize},
        xticklabels={1e-2,1e-3,1e-4,1e-5,1e-6,1e-7},
        scaled y ticks = false,
        ymode=log,
        log basis y={10},
        ytick = {100,1000,10000,100000,1000000,10000000},
        every axis y label/.style={at={(current axis.north west)},right=9mm,above=0mm},
        every axis x label/.style={at={(current axis.right of origin)},above=6mm,anchor=north west},
    ]
    \addplot[line width=0.25mm,mark=triangle]  
        plot coordinates {
(1,	200.174	)
(2,	1107.36	)
(3,	21195.3	)
(4,	81315	)
(5,	155653	)
(6,	211883	)
    };


    \addplot[line width=0.25mm,mark=square]  
        plot coordinates {
(1,	13321.77	)
(2,	319188	)
(3,	0	)
(4,	0	)
(5,	0	)
(6,	0	)
    };
    
    \addplot[line width=0.25mm,mark=o]  
        plot coordinates {
(1,	393041	)
(2,	569453	)
(3,	787095	)
(4,	1006840	)
(5,	1190620	)
(6,	1452170	)
    };
    
    \end{axis}
\end{tikzpicture}\hspace{4mm}\vspace{0mm}%
}%
\subfloat[{\em AOL}]{
\begin{tikzpicture}[scale=1]
    \begin{axis}[
        height=\columnwidth/2.7,
        width=\columnwidth/1.9,
        ylabel={\em query time (ms)},
        xlabel=$\epsilon$,
        xmin=0.5, xmax=6.5,
        ymin=10000, ymax=10000000,
        xtick={1,2,3,4,5,6},
        xticklabel style = {font=\scriptsize},
        yticklabel style = {font=\scriptsize},
        xticklabels={1e-2,1e-3,1e-4,1e-5,1e-6,1e-7},
        scaled y ticks = false,
        ymode=log,
        log basis y={10},
        ytick = {10000,100000,1000000,10000000},
        every axis y label/.style={at={(current axis.north west)},right=9mm,above=0mm},
        every axis x label/.style={at={(current axis.right of origin)},above=21mm,anchor=north west},
    ]
    \addplot[line width=0.25mm,mark=triangle]  
        plot coordinates {
(1,	10983.7	)
(2,	64181	)
(3,	113208	)
(4,	202380	)
(5,	354247	)
(6,	587610	)
    };

    \addplot[line width=0.25mm,mark=square]  
        plot coordinates {
(1,	22876.91	)
(2,	1276241	)
(3,	0	)
(4,	0	)
(5,	0	)
(6,	0	)
    };
    
    \addplot[line width=0.25mm,mark=o]  
        plot coordinates {
(1,	548999	)
(2,	799775	)
(3,	1047150	)
(4,	1223300	)
(5,	1484160	)
(6,	1822800	)
    };
    
    \end{axis}
\end{tikzpicture}\hspace{4mm}\vspace{0mm}%
}%
\vspace{-2mm}
\end{small}
\caption{Query time with varying $\epsilon$.} \label{fig:time}
\vspace{0mm}
\end{figure}
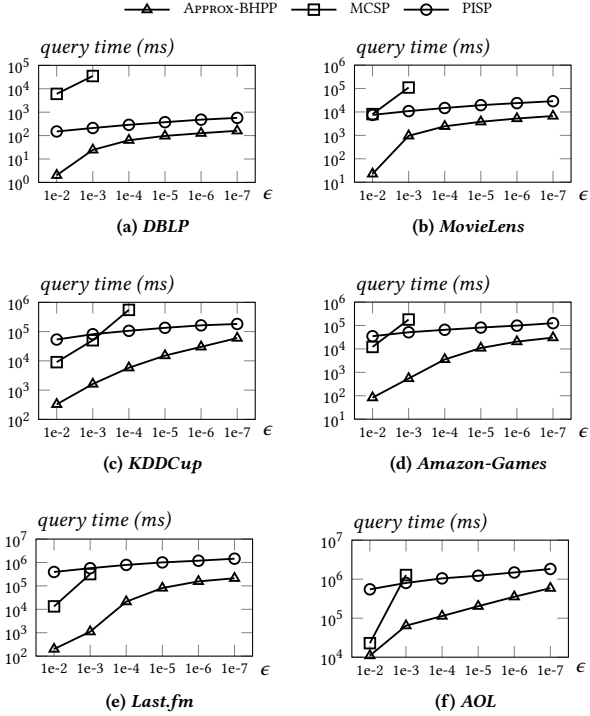

\subsection{Item Recommendation}\label{sec:ir}
\header
{\bf Settings.} Given a user-item graph $G$ and a user $v$, the goal of item recommendation is to recommend a list of interesting items to $u$. To assess the recommendation performance of BHPP and other similarity measures compared in Section \ref{sec:qr}, we adopt the evaluation methodology described in \cite{bellogin2011precision}. Specifically, we split the original graph $G$ into a training set $\overline{G}$ and a test set $\underline{G}$ such that for each user $80\%$ of its edges are in $\overline{G}$ and the remaining edges are in $\underline{G}$. After that, for each user $v$, we generate a list $L_v$ containing user-item pairs $(v,u_i)$ in $\underline{G}$ as well as user-item pairs $(v,u_j)$ satisfying $u_j\in \underline{G}$ and $(v,u_j)\notin G$. Following previous work \cite{nguyen2015evaluation}, we find a set ${S}(u_i)$ containing the items $u_j$ that are most similar to a given item $u_i$ according to the similarity $s(u_i,u_j)$ computed on $\overline{G}$. 
\begin{equation}\label{eq:pred-s}
\textstyle p(v,u_j)=\frac{\sum_{u_j\in {S}(u_i)\cup {N}(v)}{s(u_i,u_j)\cdot w(v,u_j)}}{\sum_{u_j\in {S}(u_i)\cup {N}(v)}{s(u_i,u_j)}}.
\end{equation}
Based on the similarities, the prediction score $p(v,u_i)$ for each user-item pair $(v,u_i)$ in $L_v$ is then calculated by Eq. \eqref{eq:pred-s} \cite{sarwar2001item}. 
Let the items with top-$k$ largest prediction scores in $L_v$ be the recommendations for user $v$, and the items that are connected with $v$ in $\underline{G}$ be the ground-truth list. The recommendation quality can be measured using two classical metrics: precision ($precision@k$ ) and recall ($recall@k$).
The former is defined as the fraction of the top-$k$ predicted items relevant to user $v$ and the latter as the fraction of relevant items from $\underline{G}$ that appear in the $k$ recommended items.

\header
{\bf Results.} Table \ref{tab:ir-acc-10} reports the recommendation performance of BHPP and eight competing similarity measures on four user-item graphs: {\em DBLP}, {\em MovieLens}, {\em Last.fm}, and {\em Amazon-Games} with the typical setting $k=10$ \cite{deshpande2004item}. BHPP consistently yields the best performance on four datasets. More specifically, on {\em DBLP}, BHPP achieves an impressive improvement of at least $0.9\%$ (resp. $0.7\%$) in precision (resp. recall) over existing similarity measures, and a significant margin of at least $6.3\%$ (resp. $4.4\%$) in terms of precision (resp. recall) on {\em MovieLens} compared to the state-of-the-art results. On large graphs {\em Last.fm} and {\em Amazon-Games}, BHPP still outperforms the competing similarity measures, by a considerable gain of about $0.9\%$ (resp. $1.5\%$) and a substantial margin of around $5.4\%$ (resp. $3.7\%$) in terms of precision (resp. recall). The results on $k=5$ are quantitatively similar, and, thus, are included in Appendix.

\subsection{Query Efficiency}\label{sec:time}
In this section, we evaluate the efficiency of our \algo against two baseline solutions for $\epsilon$-approximate BHPP queries mentioned in Section \ref{sec:baseline}, dubbed as \mcsp and \pisp, respectively.

\header
{\bf Settings.} We test the query performance on six datasets including {\em DBLP}, {\em MovieLens}, {\em KDDCup}, {\em Last.fm}, {\em Amazin-Games}, and {\em AOL}. For each dataset, we choose $100$ query nodes from $U$ uniformly at random. Following previous work \cite{deng2009generalized,epasto2014reduce}, we set $\alpha=0.15$. For the \mcsp approach, we set the failure probability to $p_f=10^{-6}$. In \algo, parameters $\lambda,\epsilon_b$ are computed according to Eq. \eqref{eq:lambda} and Eq. \eqref{eq:epsilon-b} in the pre-processing step, respectively. The error threshold $\epsilon$ is varied in the range $\{10^{-2},10^{-3},10^{-4},10^{-5},10^{-6},10^{-7}\}$. We report the average query times (measured in wall-clock time) of each method on each dataset with various $\epsilon$. 

\header
{\bf Results.} Figure \ref{fig:time} plots the average query time required by each method on each dataset, when varying $\epsilon$ from $10^{-2}$ to $10^{-7}$. Note that the $y$-axis is in log-scale and the measurement unit for query time is millisecond (ms). We exclude a method if it fails to answer the query within one hour on average. We
observe that \algo outperforms all competitors, often by an order of magnitude when $10^{-2}\le \epsilon \le 10^{-5}$. Notably, when $\epsilon=10^{-2}$, \algo is two orders of magnitude faster than the best competitors on all datasets except {\em AOL}, where it has comparable performance to \mcsp. Note that \mcsp is unable to answer high-precision (\eg $\epsilon\in\{10^{-6},10^{-7}\}$) queries efficiently, whereas \algo achieves $3$-$7.6\times$ speedup compared to \pisp, demonstrating the power of our optimization techniques developed in Section \ref{sec:algo}.

\section{Conclusion}\label{sec:conclude}
In this paper, we formulate similarity search over bipartite graphs as the novel problem of approximate BHPP queries, and present an efficient and theoretically-grounded solution, \algo. Our \algo combines deterministic graph traversal and matrix power iterations in an optimized way, thereby overcoming the deficiencies of both. In addition, \algo offers rigorous theoretical guarantees in terms of accuracy and time complexity, and achieves significant gain in efficiency due to several technical optimizations. Our experimental results show the superior effectiveness of BHPP compared to existing similarity measures, and demonstrate that our \algo outperforms baseline solutions by up to orders of magnitude in terms of computational time. For future work, we intend to study how to extend \algo to handle temporal/dynamic bipartite graphs.

\balance
\bibliographystyle{ACM-Reference-Format}
\bibliography{main}


\begin{thebibliography}{88}


\ifx \showCODEN    \undefined \def \showCODEN     #1{\unskip}     \fi
\ifx \showDOI      \undefined \def \showDOI       #1{#1}\fi
\ifx \showISBNx    \undefined \def \showISBNx     #1{\unskip}     \fi
\ifx \showISBNxiii \undefined \def \showISBNxiii  #1{\unskip}     \fi
\ifx \showISSN     \undefined \def \showISSN      #1{\unskip}     \fi
\ifx \showLCCN     \undefined \def \showLCCN      #1{\unskip}     \fi
\ifx \shownote     \undefined \def \shownote      #1{#1}          \fi
\ifx \showarticletitle \undefined \def \showarticletitle #1{#1}   \fi
\ifx \showURL      \undefined \def \showURL       {\relax}        \fi
\providecommand\bibfield[2]{#2}
\providecommand\bibinfo[2]{#2}
\providecommand\natexlab[1]{#1}
\providecommand\showeprint[2][]{arXiv:#2}

\bibitem[m1-(2003)]%
        {m1-1m}
 \bibinfo{year}{2003}\natexlab{}.
\newblock \bibinfo{title}{MovieLens 1M Dataset}.
\newblock
\newblock
\urldef\tempurl%
\url{https://grouplens.org/datasets/movielens}
\showURL{%
Retrieved Oct, 2021 from \tempurl}


\bibitem[aol(2006)]%
        {aol}
 \bibinfo{year}{2006}\natexlab{}.
\newblock \bibinfo{title}{AOL Query Logs}.
\newblock
\newblock
\urldef\tempurl%
\url{http://www.cim.mcgill.ca/~dudek/206/Logs/AOL-user-ct-collection}
\showURL{%
Retrieved Oct, 2021 from \tempurl}


\bibitem[las(2010)]%
        {lastfm}
 \bibinfo{year}{2010}\natexlab{}.
\newblock \bibinfo{title}{Last.fm Dataset Version 1.2}.
\newblock
\newblock
\urldef\tempurl%
\url{http://ocelma.net/MusicRecommendationDataset/lastfm-360K.html}
\showURL{%
Retrieved Oct, 2021 from \tempurl}


\bibitem[kdd(2012)]%
        {kddcup}
 \bibinfo{year}{2012}\natexlab{}.
\newblock \bibinfo{title}{KDD Cup 2012, Track 2}.
\newblock
\newblock
\urldef\tempurl%
\url{https://www.kaggle.com/c/kddcup2012-track2}
\showURL{%
Retrieved Oct, 2021 from \tempurl}


\bibitem[ama(2014)]%
        {amazon-games}
 \bibinfo{year}{2014}\natexlab{}.
\newblock \bibinfo{title}{Amazon product data}.
\newblock
\newblock
\urldef\tempurl%
\url{https://jmcauley.ucsd.edu/data/amazon}
\showURL{%
Retrieved Oct, 2021 from \tempurl}


\bibitem[avi(2015)]%
        {avito}
 \bibinfo{year}{2015}\natexlab{}.
\newblock \bibinfo{title}{Avito Context Ad Clicks}.
\newblock
\newblock
\urldef\tempurl%
\url{https://www.kaggle.com/c/avito-context-ad-clicks/data}
\showURL{%
Retrieved Oct, 2021 from \tempurl}


\bibitem[Adamic and Adar(2003)]%
        {adamic2003friends}
\bibfield{author}{\bibinfo{person}{Lada~A Adamic} {and} \bibinfo{person}{Eytan Adar}.} \bibinfo{year}{2003}\natexlab{}.
\newblock \showarticletitle{Friends and neighbors on the web}.
\newblock \bibinfo{journal}{\emph{Social networks}} (\bibinfo{year}{2003}), \bibinfo{pages}{211--230}.
\newblock


\bibitem[Anastasakos et~al\mbox{.}(2009)]%
        {anastasakos2009collaborative}
\bibfield{author}{\bibinfo{person}{Tasos Anastasakos}, \bibinfo{person}{Dustin Hillard}, \bibinfo{person}{Sanjay Kshetramade}, {and} \bibinfo{person}{Hema Raghavan}.} \bibinfo{year}{2009}\natexlab{}.
\newblock \showarticletitle{A collaborative filtering approach to ad recommendation using the query-ad click graph}. In \bibinfo{booktitle}{\emph{CIKM}}. \bibinfo{pages}{1927--1930}.
\newblock


\bibitem[Andersen et~al\mbox{.}(2008)]%
        {andersen2008local}
\bibfield{author}{\bibinfo{person}{Reid Andersen}, \bibinfo{person}{Christian Borgs}, \bibinfo{person}{Jennifer Chayes}, \bibinfo{person}{John Hopcroft}, \bibinfo{person}{Vahab Mirrokni}, {and} \bibinfo{person}{Shang-Hua Teng}.} \bibinfo{year}{2008}\natexlab{}.
\newblock \showarticletitle{Local computation of pagerank contributions}.
\newblock \bibinfo{journal}{\emph{Internet Mathematics}} (\bibinfo{year}{2008}), \bibinfo{pages}{23--45}.
\newblock


\bibitem[Andersen et~al\mbox{.}(2006)]%
        {andersen2006local}
\bibfield{author}{\bibinfo{person}{Reid Andersen}, \bibinfo{person}{Fan Chung}, {and} \bibinfo{person}{Kevin Lang}.} \bibinfo{year}{2006}\natexlab{}.
\newblock \showarticletitle{Local graph partitioning using pagerank vectors}. In \bibinfo{booktitle}{\emph{FOCS}}. \bibinfo{pages}{475--486}.
\newblock


\bibitem[Antonellis et~al\mbox{.}(2008)]%
        {antonellis2008simrank++}
\bibfield{author}{\bibinfo{person}{Ioannis Antonellis}, \bibinfo{person}{Hector~Garcia Molina}, {and} \bibinfo{person}{Chi~Chao Chang}.} \bibinfo{year}{2008}\natexlab{}.
\newblock \showarticletitle{Simrank++: Query Rewriting through Link Analysis of the Click Graph}. In \bibinfo{booktitle}{\emph{PVLDB}}. \bibinfo{pages}{408–421}.
\newblock


\bibitem[Avrachenkov et~al\mbox{.}(2007)]%
        {avrachenkov2007monte}
\bibfield{author}{\bibinfo{person}{Konstantin Avrachenkov}, \bibinfo{person}{Nelly Litvak}, \bibinfo{person}{Danil Nemirovsky}, {and} \bibinfo{person}{Natalia Osipova}.} \bibinfo{year}{2007}\natexlab{}.
\newblock \showarticletitle{Monte Carlo methods in PageRank computation: When one iteration is sufficient}.
\newblock \bibinfo{journal}{\emph{SINUM}} (\bibinfo{year}{2007}), \bibinfo{pages}{890--904}.
\newblock


\bibitem[Bahmani et~al\mbox{.}(2010)]%
        {bahmani2010fast}
\bibfield{author}{\bibinfo{person}{Bahman Bahmani}, \bibinfo{person}{Abdur Chowdhury}, {and} \bibinfo{person}{Ashish Goel}.} \bibinfo{year}{2010}\natexlab{}.
\newblock \showarticletitle{Fast Incremental and Personalized PageRank}.
\newblock \bibinfo{journal}{\emph{PVLDB}} (\bibinfo{year}{2010}).
\newblock


\bibitem[Bellogin et~al\mbox{.}(2011)]%
        {bellogin2011precision}
\bibfield{author}{\bibinfo{person}{Alejandro Bellogin}, \bibinfo{person}{Pablo Castells}, {and} \bibinfo{person}{Ivan Cantador}.} \bibinfo{year}{2011}\natexlab{}.
\newblock \showarticletitle{Precision-oriented evaluation of recommender systems: an algorithmic comparison}. In \bibinfo{booktitle}{\emph{RecSys}}. \bibinfo{pages}{333--336}.
\newblock


\bibitem[Berkhin(2005)]%
        {berkhin2005survey}
\bibfield{author}{\bibinfo{person}{Pavel Berkhin}.} \bibinfo{year}{2005}\natexlab{}.
\newblock \showarticletitle{A survey on PageRank computing}.
\newblock \bibinfo{journal}{\emph{Internet mathematics}} (\bibinfo{year}{2005}), \bibinfo{pages}{73--120}.
\newblock


\bibitem[Berkhin(2006)]%
        {berkhin2006bookmark}
\bibfield{author}{\bibinfo{person}{Pavel Berkhin}.} \bibinfo{year}{2006}\natexlab{}.
\newblock \showarticletitle{Bookmark-coloring algorithm for personalized pagerank computing}.
\newblock \bibinfo{journal}{\emph{Internet Mathematics}} (\bibinfo{year}{2006}), \bibinfo{pages}{41--62}.
\newblock


\bibitem[Celma(2010)]%
        {Celma:Springer2010}
\bibfield{author}{\bibinfo{person}{O. Celma}.} \bibinfo{year}{2010}\natexlab{}.
\newblock \bibinfo{booktitle}{\emph{{Music Recommendation and Discovery in the Long Tail}}}.
\newblock \bibinfo{publisher}{Springer}.
\newblock


\bibitem[Chakrabarti(2007)]%
        {chakrabarti2007dynamic}
\bibfield{author}{\bibinfo{person}{Soumen Chakrabarti}.} \bibinfo{year}{2007}\natexlab{}.
\newblock \showarticletitle{Dynamic personalized pagerank in entity-relation graphs}. In \bibinfo{booktitle}{\emph{WWW}}. \bibinfo{pages}{571--580}.
\newblock


\bibitem[Deng et~al\mbox{.}(2009)]%
        {deng2009generalized}
\bibfield{author}{\bibinfo{person}{Hongbo Deng}, \bibinfo{person}{Michael~R Lyu}, {and} \bibinfo{person}{Irwin King}.} \bibinfo{year}{2009}\natexlab{}.
\newblock \showarticletitle{A generalized co-hits algorithm and its application to bipartite graphs}. In \bibinfo{booktitle}{\emph{SIGKDD}}. \bibinfo{pages}{239--248}.
\newblock


\bibitem[DeSantis et~al\mbox{.}(2011)]%
        {desantis2011simrank}
\bibfield{author}{\bibinfo{person}{Todd~Z DeSantis}, \bibinfo{person}{Keith Keller}, \bibinfo{person}{Ulas Karaoz}, \bibinfo{person}{Alexander~V Alekseyenko}, \bibinfo{person}{Navjeet~NS Singh}, \bibinfo{person}{Eoin~L Brodie}, \bibinfo{person}{Zhiheng Pei}, \bibinfo{person}{Gary~L Andersen}, {and} \bibinfo{person}{Niels Larsen}.} \bibinfo{year}{2011}\natexlab{}.
\newblock \showarticletitle{Simrank: Rapid and sensitive general-purpose k-mer search tool}.
\newblock \bibinfo{journal}{\emph{BMC ecology}} (\bibinfo{year}{2011}), \bibinfo{pages}{1--8}.
\newblock


\bibitem[Deshpande and Karypis(2004)]%
        {deshpande2004item}
\bibfield{author}{\bibinfo{person}{Mukund Deshpande} {and} \bibinfo{person}{George Karypis}.} \bibinfo{year}{2004}\natexlab{}.
\newblock \showarticletitle{Item-based top-n recommendation algorithms}.
\newblock \bibinfo{journal}{\emph{TOIS}} (\bibinfo{year}{2004}), \bibinfo{pages}{143--177}.
\newblock


\bibitem[Dey et~al\mbox{.}(2020)]%
        {dey2020p}
\bibfield{author}{\bibinfo{person}{Prasenjit Dey}, \bibinfo{person}{Kunal Goel}, {and} \bibinfo{person}{Rahul Agrawal}.} \bibinfo{year}{2020}\natexlab{}.
\newblock \showarticletitle{P-Simrank: Extending Simrank to Scale-free bipartite networks}. In \bibinfo{booktitle}{\emph{The Web Conference}}. \bibinfo{pages}{3084--3090}.
\newblock


\bibitem[Epasto et~al\mbox{.}(2014)]%
        {epasto2014reduce}
\bibfield{author}{\bibinfo{person}{Alessandro Epasto}, \bibinfo{person}{Jon Feldman}, \bibinfo{person}{Silvio Lattanzi}, \bibinfo{person}{Stefano Leonardi}, {and} \bibinfo{person}{Vahab Mirrokni}.} \bibinfo{year}{2014}\natexlab{}.
\newblock \showarticletitle{Reduce and aggregate: similarity ranking in multi-categorical bipartite graphs}. In \bibinfo{booktitle}{\emph{WWW}}. \bibinfo{pages}{349--360}.
\newblock


\bibitem[Fogaras et~al\mbox{.}(2005)]%
        {fogaras2005towards}
\bibfield{author}{\bibinfo{person}{D{\'a}niel Fogaras}, \bibinfo{person}{Bal{\'a}zs R{\'a}cz}, \bibinfo{person}{K{\'a}roly Csalog{\'a}ny}, {and} \bibinfo{person}{Tam{\'a}s Sarl{\'o}s}.} \bibinfo{year}{2005}\natexlab{}.
\newblock \showarticletitle{Towards scaling fully personalized pagerank: Algorithms, lower bounds, and experiments}.
\newblock \bibinfo{journal}{\emph{Internet Mathematics}} (\bibinfo{year}{2005}), \bibinfo{pages}{333--358}.
\newblock


\bibitem[Fujiwara et~al\mbox{.}(2012)]%
        {fujiwara2012fast}
\bibfield{author}{\bibinfo{person}{Yasuhiro Fujiwara}, \bibinfo{person}{Makoto Nakatsuji}, \bibinfo{person}{Makoto Onizuka}, {and} \bibinfo{person}{Masaru Kitsuregawa}.} \bibinfo{year}{2012}\natexlab{}.
\newblock \showarticletitle{Fast and exact top-k search for random walk with restart}.
\newblock \bibinfo{journal}{\emph{PVLDB}} (\bibinfo{year}{2012}), \bibinfo{pages}{442--453}.
\newblock


\bibitem[Gao et~al\mbox{.}(2018)]%
        {BiNE18}
\bibfield{author}{\bibinfo{person}{Ming Gao}, \bibinfo{person}{Leihui Chen}, \bibinfo{person}{Xiangnan He}, {and} \bibinfo{person}{Aoying Zhou}.} \bibinfo{year}{2018}\natexlab{}.
\newblock \showarticletitle{BiNE: Bipartite Network Embedding}. \bibinfo{pages}{715--724}.
\newblock


\bibitem[Guo et~al\mbox{.}(2017a)]%
        {guo2017distributed}
\bibfield{author}{\bibinfo{person}{Tao Guo}, \bibinfo{person}{Xin Cao}, \bibinfo{person}{Gao Cong}, \bibinfo{person}{Jiaheng Lu}, {and} \bibinfo{person}{Xuemin Lin}.} \bibinfo{year}{2017}\natexlab{a}.
\newblock \showarticletitle{Distributed algorithms on exact personalized pagerank}. In \bibinfo{booktitle}{\emph{SIGMOD}}. \bibinfo{pages}{479--494}.
\newblock


\bibitem[Guo et~al\mbox{.}(2017b)]%
        {guo2017parallel}
\bibfield{author}{\bibinfo{person}{Wentian Guo}, \bibinfo{person}{Yuchen Li}, \bibinfo{person}{Mo Sha}, {and} \bibinfo{person}{Kian-Lee Tan}.} \bibinfo{year}{2017}\natexlab{b}.
\newblock \showarticletitle{Parallel personalized pagerank on dynamic graphs}.
\newblock \bibinfo{journal}{\emph{PVLDB}} (\bibinfo{year}{2017}), \bibinfo{pages}{93--106}.
\newblock


\bibitem[Harper and Konstan(2015)]%
        {harper2015movielens}
\bibfield{author}{\bibinfo{person}{F~Maxwell Harper} {and} \bibinfo{person}{Joseph~A Konstan}.} \bibinfo{year}{2015}\natexlab{}.
\newblock \showarticletitle{The movielens datasets: History and context}.
\newblock \bibinfo{journal}{\emph{TIIS}} \bibinfo{volume}{5}, \bibinfo{number}{4} (\bibinfo{year}{2015}), \bibinfo{pages}{1--19}.
\newblock


\bibitem[Haveliwala(2002)]%
        {haveliwala2002topic}
\bibfield{author}{\bibinfo{person}{Taher~H Haveliwala}.} \bibinfo{year}{2002}\natexlab{}.
\newblock \showarticletitle{Topic-sensitive PageRank}. In \bibinfo{booktitle}{\emph{WWW}}.
\newblock


\bibitem[He and McAuley(2016a)]%
        {he2016ups}
\bibfield{author}{\bibinfo{person}{Ruining He} {and} \bibinfo{person}{Julian McAuley}.} \bibinfo{year}{2016}\natexlab{a}.
\newblock \showarticletitle{Ups and downs: Modeling the visual evolution of fashion trends with one-class collaborative filtering}. In \bibinfo{booktitle}{\emph{The WebConf}}. \bibinfo{pages}{507--517}.
\newblock


\bibitem[He and McAuley(2016b)]%
        {he2016vbpr}
\bibfield{author}{\bibinfo{person}{Ruining He} {and} \bibinfo{person}{Julian McAuley}.} \bibinfo{year}{2016}\natexlab{b}.
\newblock \showarticletitle{VBPR: visual Bayesian Personalized Ranking from implicit feedback}. In \bibinfo{booktitle}{\emph{AAAI}}. \bibinfo{pages}{144--150}.
\newblock


\bibitem[Herlocker et~al\mbox{.}(2004)]%
        {herlocker2004evaluating}
\bibfield{author}{\bibinfo{person}{Jonathan~L Herlocker}, \bibinfo{person}{Joseph~A Konstan}, \bibinfo{person}{Loren~G Terveen}, {and} \bibinfo{person}{John~T Riedl}.} \bibinfo{year}{2004}\natexlab{}.
\newblock \showarticletitle{Evaluating collaborative filtering recommender systems}.
\newblock \bibinfo{journal}{\emph{TOIS}} \bibinfo{volume}{22}, \bibinfo{number}{1} (\bibinfo{year}{2004}), \bibinfo{pages}{5--53}.
\newblock


\bibitem[Hou et~al\mbox{.}(2021)]%
        {hou2021massively}
\bibfield{author}{\bibinfo{person}{Guanhao Hou}, \bibinfo{person}{Xingguang Chen}, \bibinfo{person}{Sibo Wang}, {and} \bibinfo{person}{Zhewei Wei}.} \bibinfo{year}{2021}\natexlab{}.
\newblock \showarticletitle{Massively Parallel Algorithms for Personalized PageRank}.
\newblock \bibinfo{journal}{\emph{PVLDB}} (\bibinfo{year}{2021}), \bibinfo{pages}{1668--1680}.
\newblock


\bibitem[Jaccard(1912)]%
        {jaccard1912distribution}
\bibfield{author}{\bibinfo{person}{Paul Jaccard}.} \bibinfo{year}{1912}\natexlab{}.
\newblock \showarticletitle{The distribution of the flora in the alpine zone. 1}.
\newblock \bibinfo{journal}{\emph{New phytologist}} (\bibinfo{year}{1912}), \bibinfo{pages}{37--50}.
\newblock


\bibitem[J{\"a}rvelin and Kek{\"a}l{\"a}inen(2017)]%
        {jarvelin2017ir}
\bibfield{author}{\bibinfo{person}{Kalervo J{\"a}rvelin} {and} \bibinfo{person}{Jaana Kek{\"a}l{\"a}inen}.} \bibinfo{year}{2017}\natexlab{}.
\newblock \showarticletitle{IR evaluation methods for retrieving highly relevant documents}. In \bibinfo{booktitle}{\emph{SIGIR}}. \bibinfo{pages}{243--250}.
\newblock


\bibitem[Jeh and Widom(2002)]%
        {jeh2002simrank}
\bibfield{author}{\bibinfo{person}{Glen Jeh} {and} \bibinfo{person}{Jennifer Widom}.} \bibinfo{year}{2002}\natexlab{}.
\newblock \showarticletitle{Simrank: a measure of structural-context similarity}. In \bibinfo{booktitle}{\emph{SIGKDD}}. \bibinfo{pages}{538--543}.
\newblock


\bibitem[Jeh and Widom(2003)]%
        {jeh2003scaling}
\bibfield{author}{\bibinfo{person}{Glen Jeh} {and} \bibinfo{person}{Jennifer Widom}.} \bibinfo{year}{2003}\natexlab{}.
\newblock \showarticletitle{Scaling personalized web search}. In \bibinfo{booktitle}{\emph{WWW}}. \bibinfo{pages}{271--279}.
\newblock


\bibitem[Jung et~al\mbox{.}(2017)]%
        {jung2017bepi}
\bibfield{author}{\bibinfo{person}{Jinhong Jung}, \bibinfo{person}{Namyong Park}, \bibinfo{person}{Sael Lee}, {and} \bibinfo{person}{U Kang}.} \bibinfo{year}{2017}\natexlab{}.
\newblock \showarticletitle{Bepi: Fast and memory-efficient method for billion-scale random walk with restart}. In \bibinfo{booktitle}{\emph{SIGMOD}}. \bibinfo{pages}{789--804}.
\newblock


\bibitem[Kamvar et~al\mbox{.}(2003)]%
        {kamvar2003extrapolation}
\bibfield{author}{\bibinfo{person}{Sepandar~D Kamvar}, \bibinfo{person}{Taher~H Haveliwala}, \bibinfo{person}{Christopher~D Manning}, {and} \bibinfo{person}{Gene~H Golub}.} \bibinfo{year}{2003}\natexlab{}.
\newblock \showarticletitle{Extrapolation methods for accelerating PageRank computations}. In \bibinfo{booktitle}{\emph{WWW}}. \bibinfo{pages}{261--270}.
\newblock


\bibitem[Katz(1953)]%
        {katz1953new}
\bibfield{author}{\bibinfo{person}{Leo Katz}.} \bibinfo{year}{1953}\natexlab{}.
\newblock \showarticletitle{A new status index derived from sociometric analysis}.
\newblock \bibinfo{journal}{\emph{Psychometrika}} (\bibinfo{year}{1953}), \bibinfo{pages}{39--43}.
\newblock


\bibitem[Kleinberg et~al\mbox{.}(1998)]%
        {kleinberg1998authoritative}
\bibfield{author}{\bibinfo{person}{Jon~M Kleinberg} {et~al\mbox{.}}} \bibinfo{year}{1998}\natexlab{}.
\newblock \showarticletitle{Authoritative sources in a hyperlinked environment.}. In \bibinfo{booktitle}{\emph{SODA}}. \bibinfo{pages}{668--677}.
\newblock


\bibitem[Koren(2008)]%
        {koren2008factorization}
\bibfield{author}{\bibinfo{person}{Yehuda Koren}.} \bibinfo{year}{2008}\natexlab{}.
\newblock \showarticletitle{Factorization meets the neighborhood: a multifaceted collaborative filtering model}. In \bibinfo{booktitle}{\emph{SIGKDD}}. \bibinfo{pages}{426--434}.
\newblock


\bibitem[Li et~al\mbox{.}(2013)]%
        {li2013mapreduce}
\bibfield{author}{\bibinfo{person}{Lina Li}, \bibinfo{person}{Cuiping Li}, \bibinfo{person}{Hong Chen}, {and} \bibinfo{person}{Xiaoyong Du}.} \bibinfo{year}{2013}\natexlab{}.
\newblock \showarticletitle{Mapreduce-based SimRank computation and its application in social recommender system}. In \bibinfo{booktitle}{\emph{BigData Congress}}. \bibinfo{pages}{133--140}.
\newblock


\bibitem[Lin et~al\mbox{.}(2020)]%
        {lin2020index}
\bibfield{author}{\bibinfo{person}{Dandan Lin}, \bibinfo{person}{Raymond Chi-Wing Wong}, \bibinfo{person}{Min Xie}, {and} \bibinfo{person}{Victor~Junqiu Wei}.} \bibinfo{year}{2020}\natexlab{}.
\newblock \showarticletitle{Index-Free Approach with Theoretical Guarantee for Efficient Random Walk with Restart Query}. In \bibinfo{booktitle}{\emph{ICDE}}. \bibinfo{pages}{913--924}.
\newblock


\bibitem[Lin(2019)]%
        {lin2019distributed}
\bibfield{author}{\bibinfo{person}{Wenqing Lin}.} \bibinfo{year}{2019}\natexlab{}.
\newblock \showarticletitle{Distributed algorithms for fully personalized pagerank on large graphs}. In \bibinfo{booktitle}{\emph{WWW}}. \bibinfo{pages}{1084--1094}.
\newblock


\bibitem[Liu et~al\mbox{.}(2016)]%
        {liu2016powerwalk}
\bibfield{author}{\bibinfo{person}{Qin Liu}, \bibinfo{person}{Zhenguo Li}, \bibinfo{person}{John~CS Lui}, {and} \bibinfo{person}{Jiefeng Cheng}.} \bibinfo{year}{2016}\natexlab{}.
\newblock \showarticletitle{Powerwalk: Scalable personalized pagerank via random walks with vertex-centric decomposition}. In \bibinfo{booktitle}{\emph{CIKM}}. \bibinfo{pages}{195--204}.
\newblock


\bibitem[Lofgren et~al\mbox{.}(2015)]%
        {lofgren2015bidirectional}
\bibfield{author}{\bibinfo{person}{Peter Lofgren}, \bibinfo{person}{Siddhartha Banerjee}, {and} \bibinfo{person}{Ashish Goel}.} \bibinfo{year}{2015}\natexlab{}.
\newblock \showarticletitle{Bidirectional PageRank Estimation: From Average-Case to Worst-Case}. In \bibinfo{booktitle}{\emph{WAW}}. \bibinfo{pages}{164--176}.
\newblock


\bibitem[Lofgren et~al\mbox{.}(2016)]%
        {lofgren2016personalized}
\bibfield{author}{\bibinfo{person}{Peter Lofgren}, \bibinfo{person}{Siddhartha Banerjee}, {and} \bibinfo{person}{Ashish Goel}.} \bibinfo{year}{2016}\natexlab{}.
\newblock \showarticletitle{Personalized pagerank estimation and search: A bidirectional approach}. In \bibinfo{booktitle}{\emph{WSDM}}. \bibinfo{pages}{163--172}.
\newblock


\bibitem[Lofgren and Goel(2013)]%
        {lofgren2013personalized}
\bibfield{author}{\bibinfo{person}{Peter Lofgren} {and} \bibinfo{person}{Ashish Goel}.} \bibinfo{year}{2013}\natexlab{}.
\newblock \showarticletitle{Personalized pagerank to a target node}.
\newblock \bibinfo{journal}{\emph{arXiv preprint arXiv:1304.4658}} (\bibinfo{year}{2013}).
\newblock


\bibitem[Lofgren et~al\mbox{.}(2014)]%
        {lofgren2014fast}
\bibfield{author}{\bibinfo{person}{Peter~A Lofgren}, \bibinfo{person}{Siddhartha Banerjee}, \bibinfo{person}{Ashish Goel}, {and} \bibinfo{person}{C Seshadhri}.} \bibinfo{year}{2014}\natexlab{}.
\newblock \showarticletitle{Fast-ppr: Scaling personalized pagerank estimation for large graphs}. In \bibinfo{booktitle}{\emph{SIGKDD}}. \bibinfo{pages}{1436--1445}.
\newblock


\bibitem[Maehara et~al\mbox{.}(2014)]%
        {maehara2014computing}
\bibfield{author}{\bibinfo{person}{Takanori Maehara}, \bibinfo{person}{Takuya Akiba}, \bibinfo{person}{Yoichi Iwata}, {and} \bibinfo{person}{Ken-ichi Kawarabayashi}.} \bibinfo{year}{2014}\natexlab{}.
\newblock \showarticletitle{Computing personalized pagerank quickly by exploiting graph structures}.
\newblock \bibinfo{journal}{\emph{PVLDB}} (\bibinfo{year}{2014}), \bibinfo{pages}{1023--1034}.
\newblock


\bibitem[McAuley et~al\mbox{.}(2015)]%
        {mcauley2015image}
\bibfield{author}{\bibinfo{person}{Julian McAuley}, \bibinfo{person}{Christopher Targett}, \bibinfo{person}{Qinfeng Shi}, {and} \bibinfo{person}{Anton Van Den~Hengel}.} \bibinfo{year}{2015}\natexlab{}.
\newblock \showarticletitle{Image-based recommendations on styles and substitutes}. In \bibinfo{booktitle}{\emph{SIGIR}}. \bibinfo{pages}{43--52}.
\newblock


\bibitem[Mei et~al\mbox{.}(2008)]%
        {mei2008query}
\bibfield{author}{\bibinfo{person}{Qiaozhu Mei}, \bibinfo{person}{Dengyong Zhou}, {and} \bibinfo{person}{Kenneth Church}.} \bibinfo{year}{2008}\natexlab{}.
\newblock \showarticletitle{Query suggestion using hitting time}. In \bibinfo{booktitle}{\emph{CIKM}}. \bibinfo{pages}{469--478}.
\newblock


\bibitem[Nguyen et~al\mbox{.}(2015)]%
        {nguyen2015evaluation}
\bibfield{author}{\bibinfo{person}{Phuong Nguyen}, \bibinfo{person}{Paolo Tomeo}, \bibinfo{person}{Tommaso Di~Noia}, {and} \bibinfo{person}{Eugenio Di~Sciascio}.} \bibinfo{year}{2015}\natexlab{}.
\newblock \showarticletitle{An evaluation of SimRank and Personalized PageRank to build a recommender system for the Web of Data}. In \bibinfo{booktitle}{\emph{WWW}}. \bibinfo{pages}{1477--1482}.
\newblock


\bibitem[Ohsaka et~al\mbox{.}(2015)]%
        {ohsaka2015efficient}
\bibfield{author}{\bibinfo{person}{Naoto Ohsaka}, \bibinfo{person}{Takanori Maehara}, {and} \bibinfo{person}{Ken-ichi Kawarabayashi}.} \bibinfo{year}{2015}\natexlab{}.
\newblock \showarticletitle{Efficient pagerank tracking in evolving networks}. In \bibinfo{booktitle}{\emph{SIGKDD}}. \bibinfo{pages}{875--884}.
\newblock


\bibitem[Page et~al\mbox{.}(1999)]%
        {page1999pagerank}
\bibfield{author}{\bibinfo{person}{Lawrence Page}, \bibinfo{person}{Sergey Brin}, \bibinfo{person}{Rajeev Motwani}, {and} \bibinfo{person}{Terry Winograd}.} \bibinfo{year}{1999}\natexlab{}.
\newblock \bibinfo{booktitle}{\emph{The PageRank citation ranking: Bringing order to the web.}}
\newblock \bibinfo{type}{{T}echnical {R}eport}. \bibinfo{institution}{Stanford InfoLab}.
\newblock


\bibitem[Pan et~al\mbox{.}(2004)]%
        {pan2004automatic}
\bibfield{author}{\bibinfo{person}{Jia-Yu Pan}, \bibinfo{person}{Hyung-Jeong Yang}, \bibinfo{person}{Christos Faloutsos}, {and} \bibinfo{person}{Pinar Duygulu}.} \bibinfo{year}{2004}\natexlab{}.
\newblock \showarticletitle{Automatic multimedia cross-modal correlation discovery}. In \bibinfo{booktitle}{\emph{SIGKDD}}. \bibinfo{pages}{653--658}.
\newblock


\bibitem[Park et~al\mbox{.}(2019)]%
        {park2019survey}
\bibfield{author}{\bibinfo{person}{Sungchan Park}, \bibinfo{person}{Wonseok Lee}, \bibinfo{person}{Byeongseo Choe}, {and} \bibinfo{person}{Sang-Goo Lee}.} \bibinfo{year}{2019}\natexlab{}.
\newblock \showarticletitle{A survey on personalized PageRank computation algorithms}.
\newblock \bibinfo{journal}{\emph{IEEE Access}} (\bibinfo{year}{2019}), \bibinfo{pages}{163049--163062}.
\newblock


\bibitem[Pass et~al\mbox{.}(2006)]%
        {pass2006picture}
\bibfield{author}{\bibinfo{person}{Greg Pass}, \bibinfo{person}{Abdur Chowdhury}, {and} \bibinfo{person}{Cayley Torgeson}.} \bibinfo{year}{2006}\natexlab{}.
\newblock \showarticletitle{A picture of search}. In \bibinfo{booktitle}{\emph{InfoScale}}.
\newblock


\bibitem[Pavlopoulos et~al\mbox{.}(2018)]%
        {pavlopoulos2018bipartite}
\bibfield{author}{\bibinfo{person}{Georgios~A Pavlopoulos}, \bibinfo{person}{Panagiota~I Kontou}, \bibinfo{person}{Athanasia Pavlopoulou}, \bibinfo{person}{Costas Bouyioukos}, \bibinfo{person}{Evripides Markou}, {and} \bibinfo{person}{Pantelis~G Bagos}.} \bibinfo{year}{2018}\natexlab{}.
\newblock \showarticletitle{Bipartite graphs in systems biology and medicine: a survey of methods and applications}.
\newblock \bibinfo{journal}{\emph{GigaScience}} (\bibinfo{year}{2018}), \bibinfo{pages}{1--31}.
\newblock


\bibitem[Rothe and Sch{\"u}tze(2014)]%
        {rothe2014cosimrank}
\bibfield{author}{\bibinfo{person}{Sascha Rothe} {and} \bibinfo{person}{Hinrich Sch{\"u}tze}.} \bibinfo{year}{2014}\natexlab{}.
\newblock \showarticletitle{Cosimrank: A flexible \& efficient graph-theoretic similarity measure}. In \bibinfo{booktitle}{\emph{ACL}}. \bibinfo{pages}{1392--1402}.
\newblock


\bibitem[Salton et~al\mbox{.}(1993)]%
        {salton1993approaches}
\bibfield{author}{\bibinfo{person}{Gerard Salton}, \bibinfo{person}{James Allan}, {and} \bibinfo{person}{Chris Buckley}.} \bibinfo{year}{1993}\natexlab{}.
\newblock \showarticletitle{Approaches to passage retrieval in full text information systems}. In \bibinfo{booktitle}{\emph{SIGIR}}. \bibinfo{pages}{49--58}.
\newblock


\bibitem[Sarkar and Moore(2010)]%
        {sarkar2010fast}
\bibfield{author}{\bibinfo{person}{Purnamrita Sarkar} {and} \bibinfo{person}{Andrew~W Moore}.} \bibinfo{year}{2010}\natexlab{}.
\newblock \showarticletitle{Fast nearest-neighbor search in disk-resident graphs}. In \bibinfo{booktitle}{\emph{SIGKDD}}. \bibinfo{pages}{513--522}.
\newblock


\bibitem[Sarl{\'o}s et~al\mbox{.}(2006)]%
        {sarlos2006randomize}
\bibfield{author}{\bibinfo{person}{Tam{\'a}s Sarl{\'o}s}, \bibinfo{person}{Adr{\'a}s~A Bencz{\'u}r}, \bibinfo{person}{K{\'a}roly Csalog{\'a}ny}, \bibinfo{person}{D{\'a}niel Fogaras}, {and} \bibinfo{person}{Bal{\'a}zs R{\'a}cz}.} \bibinfo{year}{2006}\natexlab{}.
\newblock \showarticletitle{To randomize or not to randomize: space optimal summaries for hyperlink analysis}. In \bibinfo{booktitle}{\emph{WWW}}. \bibinfo{pages}{297--306}.
\newblock


\bibitem[Sarwar et~al\mbox{.}(2001)]%
        {sarwar2001item}
\bibfield{author}{\bibinfo{person}{Badrul Sarwar}, \bibinfo{person}{George Karypis}, \bibinfo{person}{Joseph Konstan}, {and} \bibinfo{person}{John Riedl}.} \bibinfo{year}{2001}\natexlab{}.
\newblock \showarticletitle{Item-based collaborative filtering recommendation algorithms}. In \bibinfo{booktitle}{\emph{WWW}}. \bibinfo{pages}{285--295}.
\newblock


\bibitem[Shi et~al\mbox{.}(2020)]%
        {shi2020realtime}
\bibfield{author}{\bibinfo{person}{Jieming Shi}, \bibinfo{person}{Tianyuan Jin}, \bibinfo{person}{Renchi Yang}, \bibinfo{person}{Xiaokui Xiao}, {and} \bibinfo{person}{Yin Yang}.} \bibinfo{year}{2020}\natexlab{}.
\newblock \showarticletitle{Realtime index-free single source SimRank processing on web-scale graphs}.
\newblock \bibinfo{journal}{\emph{PVLDB}} \bibinfo{volume}{13}, \bibinfo{number}{7} (\bibinfo{year}{2020}), \bibinfo{pages}{966--980}.
\newblock


\bibitem[Shi et~al\mbox{.}(2019)]%
        {shi2019realtime}
\bibfield{author}{\bibinfo{person}{Jieming Shi}, \bibinfo{person}{Renchi Yang}, \bibinfo{person}{Tianyuan Jin}, \bibinfo{person}{Xiaokui Xiao}, {and} \bibinfo{person}{Yin Yang}.} \bibinfo{year}{2019}\natexlab{}.
\newblock \showarticletitle{Realtime top-k personalized pagerank over large graphs on gpus}.
\newblock \bibinfo{journal}{\emph{PVLDB}} (\bibinfo{year}{2019}), \bibinfo{pages}{15--28}.
\newblock


\bibitem[Shin et~al\mbox{.}(2015)]%
        {shin2015bear}
\bibfield{author}{\bibinfo{person}{Kijung Shin}, \bibinfo{person}{Jinhong Jung}, \bibinfo{person}{Sael Lee}, {and} \bibinfo{person}{U Kang}.} \bibinfo{year}{2015}\natexlab{}.
\newblock \showarticletitle{Bear: Block elimination approach for random walk with restart on large graphs}. In \bibinfo{booktitle}{\emph{SIGMOD}}. \bibinfo{pages}{1571--1585}.
\newblock


\bibitem[Sun et~al\mbox{.}(2005)]%
        {sun2005neighborhood}
\bibfield{author}{\bibinfo{person}{Jimeng Sun}, \bibinfo{person}{Huiming Qu}, \bibinfo{person}{Deepayan Chakrabarti}, {and} \bibinfo{person}{Christos Faloutsos}.} \bibinfo{year}{2005}\natexlab{}.
\newblock \showarticletitle{Neighborhood Formation and Anomaly Detection in Bipartite Graphs}. In \bibinfo{booktitle}{\emph{ICDM}}. \bibinfo{pages}{418--425}.
\newblock


\bibitem[Sun et~al\mbox{.}(2011)]%
        {sun2011link}
\bibfield{author}{\bibinfo{person}{Liwen Sun}, \bibinfo{person}{Reynold Cheng}, \bibinfo{person}{Xiang Li}, \bibinfo{person}{David~W Cheung}, {and} \bibinfo{person}{Jiawei Han}.} \bibinfo{year}{2011}\natexlab{}.
\newblock \showarticletitle{On link-based similarity join}.
\newblock \bibinfo{journal}{\emph{PVLDB}} (\bibinfo{year}{2011}), \bibinfo{pages}{714--725}.
\newblock


\bibitem[Tong et~al\mbox{.}(2006)]%
        {tong2006fast}
\bibfield{author}{\bibinfo{person}{Hanghang Tong}, \bibinfo{person}{Christos Faloutsos}, {and} \bibinfo{person}{Jia-Yu Pan}.} \bibinfo{year}{2006}\natexlab{}.
\newblock \showarticletitle{Fast random walk with restart and its applications}. In \bibinfo{booktitle}{\emph{ICDM}}. IEEE, \bibinfo{pages}{613--622}.
\newblock


\bibitem[Tong et~al\mbox{.}(2008)]%
        {tong2008proximity}
\bibfield{author}{\bibinfo{person}{Hanghang Tong}, \bibinfo{person}{Spiros Papadimitriou}, \bibinfo{person}{Philip~S Yu}, {and} \bibinfo{person}{Christos Faloutsos}.} \bibinfo{year}{2008}\natexlab{}.
\newblock \showarticletitle{Proximity tracking on time-evolving bipartite graphs}. In \bibinfo{booktitle}{\emph{SDM}}. \bibinfo{pages}{704--715}.
\newblock


\bibitem[Tversky(1977)]%
        {tversky1977features}
\bibfield{author}{\bibinfo{person}{Amos Tversky}.} \bibinfo{year}{1977}\natexlab{}.
\newblock \showarticletitle{Features of similarity.}
\newblock \bibinfo{journal}{\emph{Psychological review}} (\bibinfo{year}{1977}), \bibinfo{pages}{327}.
\newblock


\bibitem[Vijaymeena and Kavitha(2016)]%
        {vijaymeena2016survey}
\bibfield{author}{\bibinfo{person}{MK Vijaymeena} {and} \bibinfo{person}{K Kavitha}.} \bibinfo{year}{2016}\natexlab{}.
\newblock \showarticletitle{A survey on similarity measures in text mining}.
\newblock \bibinfo{journal}{\emph{MLAIJ}} (\bibinfo{year}{2016}), \bibinfo{pages}{19--28}.
\newblock


\bibitem[Walker(1974)]%
        {walker1974new}
\bibfield{author}{\bibinfo{person}{Alastair~J Walker}.} \bibinfo{year}{1974}\natexlab{}.
\newblock \showarticletitle{New fast method for generating discrete random numbers with arbitrary frequency distributions}.
\newblock \bibinfo{journal}{\emph{Electronics Letters}} (\bibinfo{year}{1974}), \bibinfo{pages}{127--128}.
\newblock


\bibitem[Wang et~al\mbox{.}(2020)]%
        {wang2020personalized}
\bibfield{author}{\bibinfo{person}{Hanzhi Wang}, \bibinfo{person}{Zhewei Wei}, \bibinfo{person}{Junhao Gan}, \bibinfo{person}{Sibo Wang}, {and} \bibinfo{person}{Zengfeng Huang}.} \bibinfo{year}{2020}\natexlab{}.
\newblock \showarticletitle{Personalized PageRank to a Target Node, Revisited}. In \bibinfo{booktitle}{\emph{SIGKDD}}. \bibinfo{pages}{657--667}.
\newblock


\bibitem[Wang et~al\mbox{.}(2019a)]%
        {wang2019parallelizing}
\bibfield{author}{\bibinfo{person}{Runhui Wang}, \bibinfo{person}{Sibo Wang}, {and} \bibinfo{person}{Xiaofang Zhou}.} \bibinfo{year}{2019}\natexlab{a}.
\newblock \showarticletitle{Parallelizing approximate single-source personalized pagerank queries on shared memory}.
\newblock \bibinfo{journal}{\emph{VLDBJ}} (\bibinfo{year}{2019}), \bibinfo{pages}{923--940}.
\newblock


\bibitem[Wang et~al\mbox{.}(2016)]%
        {wang2016hubppr}
\bibfield{author}{\bibinfo{person}{Sibo Wang}, \bibinfo{person}{Youze Tang}, \bibinfo{person}{Xiaokui Xiao}, \bibinfo{person}{Yin Yang}, {and} \bibinfo{person}{Zengxiang Li}.} \bibinfo{year}{2016}\natexlab{}.
\newblock \showarticletitle{Hubppr: effective indexing for approximate personalized pagerank}.
\newblock \bibinfo{journal}{\emph{PVLDB}} (\bibinfo{year}{2016}), \bibinfo{pages}{205--216}.
\newblock


\bibitem[Wang et~al\mbox{.}(2019b)]%
        {wang2019efficient}
\bibfield{author}{\bibinfo{person}{Sibo Wang}, \bibinfo{person}{Renchi Yang}, \bibinfo{person}{Runhui Wang}, \bibinfo{person}{Xiaokui Xiao}, \bibinfo{person}{Zhewei Wei}, \bibinfo{person}{Wenqing Lin}, \bibinfo{person}{Yin Yang}, {and} \bibinfo{person}{Nan Tang}.} \bibinfo{year}{2019}\natexlab{b}.
\newblock \showarticletitle{Efficient algorithms for approximate single-source personalized pagerank queries}.
\newblock \bibinfo{journal}{\emph{TODS}} (\bibinfo{year}{2019}), \bibinfo{pages}{1--37}.
\newblock


\bibitem[Wang et~al\mbox{.}(2017)]%
        {wang2017fora}
\bibfield{author}{\bibinfo{person}{Sibo Wang}, \bibinfo{person}{Renchi Yang}, \bibinfo{person}{Xiaokui Xiao}, \bibinfo{person}{Zhewei Wei}, {and} \bibinfo{person}{Yin Yang}.} \bibinfo{year}{2017}\natexlab{}.
\newblock \showarticletitle{FORA: simple and effective approximate single-source personalized pagerank}. In \bibinfo{booktitle}{\emph{SIGKDD}}. \bibinfo{pages}{505--514}.
\newblock


\bibitem[Wei et~al\mbox{.}(2018)]%
        {wei2018topppr}
\bibfield{author}{\bibinfo{person}{Zhewei Wei}, \bibinfo{person}{Xiaodong He}, \bibinfo{person}{Xiaokui Xiao}, \bibinfo{person}{Sibo Wang}, \bibinfo{person}{Shuo Shang}, {and} \bibinfo{person}{Ji-Rong Wen}.} \bibinfo{year}{2018}\natexlab{}.
\newblock \showarticletitle{Topppr: top-k personalized pagerank queries with precision guarantees on large graphs}. In \bibinfo{booktitle}{\emph{SIGMOD}}. \bibinfo{pages}{441--456}.
\newblock


\bibitem[Wu et~al\mbox{.}(2021)]%
        {wu2021unifying}
\bibfield{author}{\bibinfo{person}{Hao Wu}, \bibinfo{person}{Junhao Gan}, \bibinfo{person}{Zhewei Wei}, {and} \bibinfo{person}{Rui Zhang}.} \bibinfo{year}{2021}\natexlab{}.
\newblock \showarticletitle{Unifying the Global and Local Approaches: An Efficient Power Iteration with Forward Push}. In \bibinfo{booktitle}{\emph{SIGMOD}}.
\newblock


\bibitem[Yang and Xiao(2021)]%
        {yang2021fast}
\bibfield{author}{\bibinfo{person}{Renchi Yang} {and} \bibinfo{person}{Xiaokui Xiao}.} \bibinfo{year}{2021}\natexlab{}.
\newblock \showarticletitle{Fast Approximate All Pairwise CoSimRanks via Random Projection}. In \bibinfo{booktitle}{\emph{WISE}}. \bibinfo{pages}{438--452}.
\newblock


\bibitem[Yoon et~al\mbox{.}(2018)]%
        {yoon2018tpa}
\bibfield{author}{\bibinfo{person}{Minji Yoon}, \bibinfo{person}{Jinhong Jung}, {and} \bibinfo{person}{U Kang}.} \bibinfo{year}{2018}\natexlab{}.
\newblock \showarticletitle{Tpa: Fast, scalable, and accurate method for approximate random walk with restart on billion scale graphs}. In \bibinfo{booktitle}{\emph{ICDE}}. \bibinfo{pages}{1132--1143}.
\newblock


\bibitem[Yu and Lin(2013)]%
        {yu2013irwr}
\bibfield{author}{\bibinfo{person}{Weiren Yu} {and} \bibinfo{person}{Xuemin Lin}.} \bibinfo{year}{2013}\natexlab{}.
\newblock \showarticletitle{IRWR: incremental random walk with restart}. In \bibinfo{booktitle}{\emph{SIGIR}}. \bibinfo{pages}{1017--1020}.
\newblock


\bibitem[Zhang et~al\mbox{.}(2016)]%
        {zhang2016approximate}
\bibfield{author}{\bibinfo{person}{Hongyang Zhang}, \bibinfo{person}{Peter Lofgren}, {and} \bibinfo{person}{Ashish Goel}.} \bibinfo{year}{2016}\natexlab{}.
\newblock \showarticletitle{Approximate personalized pagerank on dynamic graphs}. In \bibinfo{booktitle}{\emph{SIGKDD}}. \bibinfo{pages}{1315--1324}.
\newblock


\bibitem[Zhu et~al\mbox{.}(2013)]%
        {zhu2013incremental}
\bibfield{author}{\bibinfo{person}{Fanwei Zhu}, \bibinfo{person}{Yuan Fang}, \bibinfo{person}{Kevin Chen-Chuan Chang}, {and} \bibinfo{person}{Jing Ying}.} \bibinfo{year}{2013}\natexlab{}.
\newblock \showarticletitle{Incremental and accuracy-aware personalized pagerank through scheduled approximation}.
\newblock \bibinfo{journal}{\emph{PVLDB}} (\bibinfo{year}{2013}), \bibinfo{pages}{481--492}.
\newblock


\end{thebibliography}

\balance
\appendix
\section{Proofs}\label{sec:proof}
\begin{table*}[ht]
\centering
\renewcommand{\arraystretch}{1.0}
\caption{Item recommendation performance ($k=5$).}\vspace{-2mm}
\begin{footnotesize}
\begin{tabular}{|c|c c|c c|c c| c c|}
\hline
\multirow{2}{*}{\bf Similarity} & \multicolumn{2}{c|}{\bf {\em DBLP}} & \multicolumn{2}{c|}{\bf {\em Movielens}} & \multicolumn{2}{c|}{\bf {\em Last.fm}} &
\multicolumn{2}{c|}{\bf {\em Amazon-Games}} \\ \cline{2-9}
& $precision@k$ & $recall@k$ & $precision@k$ & $recall@k$ & $precision@k$ & $recall@k$ &  $precision@k$ & $recall@k$ \\ 
\hline
BHPP & {\bf 0.165} & {\bf 0.115} & {\bf 0.609} & {\bf 0.22} & {\bf 0.441} & {\bf 0.163} & {\bf 0.36}  & {\bf 0.136} \\ 
\hline
HPP  & 0.15 & 0.097 & 0.291 & 0.105 & 0.416 & 0.15 &  0.28 & 0.108 \\ 
Pearson  & 0.095 & 0.064 & 0.091 & 0.031 & 0.178 & 0.067 &  0.104 & 0.039 \\ 
Jaccard  & 0.139 & 0.095 & 0.322 & 0.114 & 0.307 & 0.093 & 0.112  & 0.041 \\ 
SimRank  & 0.157 & 0.109 & 0.325 & 0.118 & 0.356 & 0.112 & 0.209  & 0.088 \\  
CoSimRank  & 0.152 & 0.102 & 0.322 & 0.108 & 0.415 & 0.152 &  0.243 & 0.098 \\  
PPR  & 0.127 & 0.098 & 0.475 & 0.17 & 0.393 & 0.145 & 0.272  & 0.104 \\ 
SimRank++  & 0.15 & 0.101 & 0.325 & 0.118 & 0.367 & 0.12 & 0.277  & 0.103 \\  
P-SimRank  & 0.15 & 0.1 & 0.32 & 0.112 &  0.343 & 0.108  & 0.226 & 0.094 \\  
\hline
\end{tabular}
\end{footnotesize}
\label{tab:ir-acc-5}
\vspace{1mm}
\end{table*}

\begin{figure}[h]
\centering
\begin{small}
\begin{tikzpicture}
    \begin{customlegend}[legend columns=5,
        legend entries={BHPP,HPP,Pearson,Jaccard,SimRank},
        area legend,
        legend style={at={(0.45,1.15)},anchor=north,draw=none,font=\scriptsize,column sep=0.15cm}]
        \addlegendimage{,fill=black}
        \addlegendimage{,pattern=crosshatch dots}
        \addlegendimage{,pattern=grid}
        \addlegendimage{,pattern=sixpointed stars}
        \addlegendimage{,pattern=vertical lines}
    \end{customlegend}
\end{tikzpicture}
\\[-\lineskip]
\vspace{-1mm}
\begin{tikzpicture}
    \begin{customlegend}[legend columns=4,
        legend entries={PPR,CoSimRank,SimRank++,P-SimRank},
        area legend,
        legend style={at={(0.45,1.15)},anchor=north,draw=none,font=\scriptsize,column sep=0.15cm}]
        \addlegendimage{,pattern=dots}
        \addlegendimage{,pattern=horizontal lines}
        \addlegendimage{,pattern=fivepointed stars}
        \addlegendimage{,pattern=north east lines}
    \end{customlegend}
\end{tikzpicture}
\subfloat[{\em Avito}]{
\begin{tikzpicture}[scale=1]
\begin{axis}[
    height=\columnwidth/2.7,
    width=\columnwidth/2.3,
    ybar=0.2pt,
    bar width=0.22cm,
    enlarge x limits=true,
    ylabel={\em NDCG}$@k$,
    xticklabel=\empty,
    ymin=0.7,
    ymax=0.92,
    every axis y label/.style={font=\scriptsize,at={(current axis.north west)},right=4.5mm,above=0mm},
    legend style={at={(0.02,0.98)},anchor=north west,cells={anchor=west},font=\tiny}
    ]

\addplot [,fill=black] coordinates {(1,0.913265) }; 
\addplot [,pattern=crosshatch dots] coordinates {(1,0.877666)}; 
\addplot [,pattern=grid] coordinates {(1,0.874191)}; 
\addplot [,pattern=sixpointed stars] coordinates {(1,0.884535)}; 
\addplot [,pattern=vertical lines] coordinates {(1,0.886295)}; 
\addplot [,pattern=horizontal lines] coordinates {(1,0.909711)}; 
\addplot [,pattern=dots] coordinates {(1,0.911930)}; 
\addplot [,pattern=fivepointed stars] coordinates {(1,0.885890)}; 
\addplot [,pattern=north east lines] coordinates {(1,0.886681)}; 
\end{axis}
\end{tikzpicture}\vspace{-2mm}\hspace{0mm}%
}%
\subfloat[{\em KDDCup}]{
\begin{tikzpicture}[scale=1]
\begin{axis}[
    height=\columnwidth/2.7,
    width=\columnwidth/2.3,
    ybar=0.2pt,
    bar width=0.22cm,
    enlarge x limits=true,
    ylabel={\em NDCG}$@k$,
    xticklabel=\empty,
    ymin=0.99,
    ymax=1.0,
    ytick={0.99,0.992,0.994,0.996,0.998,1.0},
    yticklabels={0.99,0.992,0.994,0.996,0.998,1.0},
    every axis y label/.style={font=\scriptsize,at={(current axis.north west)},right=4.5mm,above=0mm},
    legend style={at={(0.02,0.98)},anchor=north west,cells={anchor=west},font=\tiny}
    ]

\addplot [,fill=black] coordinates {(1,0.999818) }; 
\addplot [,pattern=crosshatch dots] coordinates {(1,0.998675)}; 
\addplot [,pattern=grid] coordinates {(1,0.99278)}; 
\addplot [,pattern=sixpointed stars] coordinates {(1,0.99359)}; 
\addplot [,pattern=vertical lines] coordinates {(1,0.998278)}; 
\addplot [,pattern=horizontal lines] coordinates {(1,0.998705)}; 
\addplot [,pattern=dots] coordinates {(1,0.99896)}; 
\addplot [,pattern=fivepointed stars] coordinates {(1,0.998278)}; 
\addplot [,pattern=north east lines] coordinates {(1,0.998278)}; 
\end{axis}
\end{tikzpicture}\vspace{-2mm}\hspace{0mm}%
}%
\subfloat[{\em AOL}]{
\begin{tikzpicture}[scale=1]
\begin{axis}[
    height=\columnwidth/2.7,
    width=\columnwidth/2.3,
    ybar=0.2pt,
    bar width=0.22cm,
    enlarge x limits=true,
    ylabel={\em NDCG}$@k$,
    xticklabel=\empty,
    ymin=0.992,
    ymax=1.0,
    ytick={0.992,0.994,0.996,0.998,1.0},
    yticklabels={0.992,0.994,0.996,0.998,1.0},
    every axis y label/.style={font=\scriptsize,at={(current axis.north west)},right=4.5mm,above=0mm},
    legend style={at={(0.02,0.98)},anchor=north west,cells={anchor=west},font=\tiny}
    ]

\addplot [,fill=black] coordinates {(1,0.999754) }; 
\addplot [,pattern=crosshatch dots] coordinates {(1,0.999568)}; 
\addplot [,pattern=grid] coordinates {(1,0.996838)}; 
\addplot [,pattern=sixpointed stars] coordinates {(1,0.99426)}; 
\addplot [,pattern=vertical lines] coordinates {(1,0.998083)}; 
\addplot [,pattern=horizontal lines] coordinates {(1,0.997445)}; 
\addplot [,pattern=dots] coordinates {(1,0.99765)}; 
\addplot [,pattern=fivepointed stars] coordinates {(1,0.998083)}; 
\addplot [,pattern=north east lines] coordinates {(1,0.998083)}; 
\end{axis}
\end{tikzpicture}\vspace{-2mm}\hspace{0mm}%
}%
\vspace{-2mm}
\end{small}
\caption{Query rewriting performance ($k=5$).} \label{fig:qr-5}
\vspace{-2mm}
\end{figure}
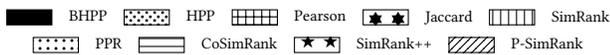
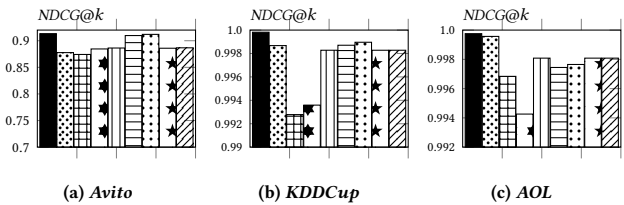

\begin{proof}[\bf Proof of Lemma \ref{lem:bwd-eq}]
We prove this lemma based on induction. First, consider the initial case, where all entries in $\bresidue_u$ are zero except $\bresidue_u(u)=1$, and $\pi_b(u_i,u)=0$ for each node $u_i\in U$. Hence, we have $\pi(u_i,u)=\pi(u_i,u)\cdot \bresidue_u(u)\ \forall{u_i\in U}$, implying Eq. \eqref{eq:bwd-eq} holds in the initial case.

Next, we assume that after $\ell$-th iterations (Lines 4-12), the approximate HPP values $\bppr^{(\ell)}(u_i,u)$ $\forall{u_i\in U}$ and residue vector $\bresidue^{(\ell)}_u$ at the end of this iteration satisfy Eq. \eqref{eq:bwd-eq}, \ie
$\textstyle\ppr(u_i,u)=\bppr^{(\ell)}(u_i,u)+\sum_{u_j\in U}{\pi(u_i,u_j)\cdot\bresidue^{(\ell)}_u(u_j)}$. Let $\bppr^{(\ell+1)}(u_i,u)$ $\forall{u_i\in U}$ be the approximate HPP values and $\bresidue^{(\ell+1)}_u$ be residue vector by the end of $(\ell+1)$-th iteration. For any node $u_i\in U$, we define $\Delta(u_i)$ as follows:
\begin{align}
\Delta(u_i)&=\bppr^{(\ell+1)}(u_i,u)-\bppr^{(\ell)}(u_i,u)\nonumber\\
&\textstyle \quad +\sum_{u_j\in U}{\left(\bresidue^{(\ell+1)}_u(u_j)-\bresidue^{(\ell)}_u(u_j)\right)\cdot \pi(u_i,u_j)}\label{eq:delta_ui}.
\end{align}
If $\Delta(u_i)=0$ holds for each $u_i\in U$, the lemma is established. According to Lines 4-12, we have 
\begin{align*}
&\textstyle\bresidue^{(\ell+1)}_u(u_j)-\bresidue^{(\ell)}_u(u_j)\\
&\textstyle=-\bresidue^{(\ell)}_u(u_j)+\sum_{\substack{u_x\in U \\ \bresidue_u(u_x)\ge \epsilon_b}}{(1-\alpha)\cdot \bresidue_u(u_x)\cdot \PM(u_j,u_x)}
\end{align*}
for node $u_j$ with $\bresidue^{(\ell)}_u(u_j)\ge\epsilon_b$. Note that  $\bresidue^{(\ell+1)}_u(u_j)-\bresidue^{(\ell)}_u(u_j)=\sum_{{u_x\in U \& \bresidue_u(u_x)\ge \epsilon_b}}{(1-\alpha)\cdot \bresidue_u(u_x)\cdot \PM(u_j,u_x)}$ holds for node $u_j$ with $\bresidue^{(\ell)}_u(u_j)<\epsilon_b$. Thus, Eq. \eqref{eq:delta_ui} becomes
\begin{align*}
\Delta(u_i)&\textstyle=\bppr^{(\ell+1)}(u_i,u)-\bppr^{(\ell)}(u_i,u)-\sum_{\substack{u_j\in U \\ \bresidue_u(u_j)\ge \epsilon_b}}{\bresidue^{(\ell)}_u(u_j) \pi(u_i,u_j)}\\
&\textstyle\quad + \sum_{u_j\in U}{\sum_{\substack{u_x\in U \\ \bresidue_u(u_x)\ge \epsilon_b}}{(1-\alpha)\cdot \PM(u_j,u_x)\cdot \pi(u_x,u_j)}}.
\end{align*}
If $\bresidue_u(u_i)< \epsilon_b$, we have 
\begin{align*}
\Delta(u_i)&\textstyle=\sum_{\substack{u_j\in U \\ \bresidue_u(u_j)\ge \epsilon_b}}{\bresidue^{(\ell)}_u(u_j)\cdot \pi(u_i,u_j)}\\
&\textstyle\quad + \sum_{u_j\in U}{\sum_{\substack{u_x\in U \\ \bresidue_u(u_x)\ge \epsilon_b}}{\bresidue^{(\ell)}_u(u_j)\cdot \pi(u_i,u_x)}}=0.
\end{align*}
Otherwise, we obtain
\begin{align*}
\Delta(u_i)&\textstyle=\alpha\cdot \bresidue_u(u_i)-\sum_{\substack{u_j\in U \\ \bresidue_u(u_j)\ge \epsilon_b}}{\bresidue^{(\ell)}_u(u_j)\cdot \pi(u_i,u_j)}\\
&\textstyle\quad + \sum_{\substack{u_x\in U, u_x\neq u_j \\ \bresidue_u(u_x)\ge \epsilon_b}}{\left(\bresidue^{(\ell)}_u(u_j)\cdot \pi(u_i,u_x)\right)}\\
&\textstyle\quad +\bresidue^{(\ell)}_u(u_i)\cdot (\pi(u_i,u_i)-\alpha)=0,
\end{align*}
which seals the proof.
\end{proof}

\begin{proof}[\bf Proof of Theorem \ref{lem:bwd-time}]
According to Lines 4-12 in Algorithm \ref{alg:bp}, in each round, each node $u_i$ involves converting $\alpha$ portion of $\bresidue_u(u_i)>\epsilon_b$ to its approximate HPP $\bppr(u_i,u)$, and distributing the remaining residue to its neighbors $N(u_i)$ as well as neighbors' neighbors. Due to $\bppr(u_i,u)\le\pi(u_i,u)$, $u_i$ requires at most $\frac{\pi(u_i,u)}{\alpha\cdot\epsilon_b}$ iterations to convert its residue into the HPP. As a result, the cost for a node $u_i$ is $\frac{\pi(u_i,u)}{\alpha\cdot\epsilon_b}\cdot\sum_{v_j\in N(u_i)}{d(v_j)}$. When considering all nodes $u_i\in U$, the total time complexity is bounded by
\begin{align*}
&\textstyle\sum_{u\in U}{\sum_{u_i\in U}}{\frac{\pi(u_i,u)}{\alpha\cdot\epsilon_b}\cdot\sum_{v_j\in N(u_i)}{d(v_j)}}\\
&\textstyle=\frac{1}{\alpha\cdot\epsilon_b}\sum_{u_i\in U}{\sum_{u\in U}{\pi(u_i,u)}\cdot\sum_{v_j\in N(u_i)}{d(v_j)}}\\
&\textstyle=\frac{1}{\alpha\cdot\epsilon_b}\sum_{u_i\in U}{\cdot\sum_{v_j\in N(u_i)}{d(v_j)}}=\frac{1}{\alpha\cdot\epsilon_b}\cdot\sum_{v_j\in V}{d(v_j)^2}.
\end{align*}
Therefore, the amortized cost of Algorithm \ref{alg:bp} is $\textstyle O\left(\frac{\sum_{v_j\in V}{d(v_j)^2}}{|U|\cdot \alpha \cdot \epsilon_b}\right)$.
\end{proof}

\begin{proof}[\bf Proof of Theorem \ref{lem:bwd}]
If \bwdp returns $\bppr(u_i,u)$ for each node $u_i\in U$ at Line 17, we have $\forall{u_i}\in U$ such that $\bresidue_{u}(u_i)\le \epsilon_b$. By Lemma \ref{lem:bwd-eq}, we obtain $\ppr(u_i,u)-\bppr(u_i,u)\le \epsilon_b$, which holds for each node $u_i\in U$.

If Algorithm \ref{alg:bwd} breaks at Line 18, \bwdp will conduct sequential push operations (Lines 19-28). This procedure can be regarded as the \bpush when $\epsilon_b$ is set to $0$. According to Lemma \ref{lem:bwd-eq}, at the end of each iteration in sequential pushes, Eq. \eqref{eq:bwd-eq} still holds. By Line 19, Algorithm \ref{alg:bwd} returns $\{\bppr(u_i,u)|u_i\in U\}$ when either (i) $\bresidue_u(u_i)\le \epsilon_b$ for each node $u_i$ in U or (ii) $\sum_{u_i\in U}{\bresidue_u(u_i)}\le\epsilon_b$. Based on Eq. \eqref{eq:bwd-eq}, we derive
\begin{gather*}
\textstyle\ppr(u_i,u)-\bppr(u_i,u) \le \epsilon_b\sum_{u_i\in U}{\pi(u_i,u_j)}=\epsilon_b\ \textrm{and}\\
\textstyle\ppr(u_i,u)-\bppr(u_i,u) \le \sum_{u_i\in U}{\bresidue_u(u_i)}\le\epsilon_b,
\end{gather*}
respectively, which complete our proof.
\end{proof}

\begin{proof}[\bf Proof of Lemma \ref{lem:asym}]
According to Eq. \eqref{eq:hpp},
\begin{equation*}
\textstyle\pi(u,u_i)=\sum_{\ell=0}^{\infty}{\alpha(1-\alpha)^{\ell}\cdot\PM^{\ell}(u,u_i)},
\end{equation*}
 which suggests that if we can prove for any $t\ge 1$,
\begin{equation}\label{eq:pmt}\textstyle
\frac{\PM^t(u,u_i)}{ws(u_i)}=\frac{\PM^t(u_i,u)}{ws(u)}, 
\end{equation}
we establishes the lemma. Next, we prove Eq. \eqref{eq:pmt} by induction. For the based case, \ie $t=1$, Eq. \eqref{eq:pmatrix} implies
\begin{align*}
\textstyle\frac{\PM(u_i,u_j)}{ws(u_j)}&\textstyle=\frac{1}{ws(u_j)}\sum_{v_l\in N(u_i)\cap N(u_j)}{\frac{w(u_i,v_l)}{ws(u_i)}\cdot \frac{w(v_l,u_j)}{ws(v_l)}}\\
&\textstyle=\frac{1}{ws(u_i)}\sum_{v_l\in N(u_i)\cap N(u_j)}{\frac{w(u_j,v_l)}{ws(u_j)}\cdot \frac{w(v_l,u_i)}{ws(v_l)}}=\frac{\PM(u_j,u_i)}{ws(u_i)}.
\end{align*}
Thus, Eq. \eqref{eq:pmt} holds when $t=1$. Assume that we have Eq. \eqref{eq:pmt} for any $u_j\in U$ when $t=\ell$. Then we consider the case for $t=\ell+1$, which satisfies
\begin{align*}
\textstyle\frac{\PM^{\ell+1}(u,u_i)}{ws(u_i)}&\textstyle=\sum_{u_j\in N(u_i)}{\frac{\PM^{\ell}(u,u_j)\cdot\PM(u_j,u_i)}{ws(u_i)}}\\
&\textstyle=\sum_{u_j\in N(u_i)}{\frac{\PM^{\ell}(u,u_j)\cdot\PM(u_i,u_j)}{ws(u_j)}}\\
&\textstyle=\sum_{u_j\in N(u_i)}{\frac{\PM^{\ell}(u_j,u)\cdot\PM(u_i,u_j)}{ws(u)}}=\frac{\PM^{\ell+1}(u_i,u)}{ws(u)}.
\end{align*}
Therefore, the lemma is proved.
\end{proof}

\begin{proof}[\bf Proof of Theorem \ref{lem:fwd}]
First, let $\fppr_u(u_i)=\bppr_u(u_i)\cdot\frac{ws(u_i)}{ws(u)}$ and $\fresidue_u(u_j)=\bresidue_u(u_j)\cdot\frac{ws(u_j)}{ws(u)}$ (Lines 13-15). By Eq. \eqref{eq:bwd2fwd}, we have
\begin{equation}\label{eq:bwd-fwd}
\textstyle\pi(u,u_i)=\textstyle\fppr(u,u_i)+\sum_{u_j\in U}{\fresidue_u(u_j)\cdot\pi(u_j,u_i)}
\end{equation}
holds for each node $u_i\in U$.

If $\forall{u_j\in U}$ $\fresidue_{u}(u_j)\le \frac{\epsilon_f}{\lambda}$, plugging Eq. \eqref{eq:lambda-ineq} into Eq. \eqref{eq:bwd-fwd}, we get for each node $u_i\in U$, 
\begin{align*}
\textstyle \pi(u,u_i)-\fppr(u,u_i)\le \epsilon_f\cdot \frac{\sum_{u_j\in U}{\pi(u_j,u_i)}}{\lambda}\le \epsilon_f.
\end{align*}

Next, consider the case where $\forall u_i\in U$ such that $\fresidue_{u}(u_i)\le \frac{\epsilon_f}{\lambda}$. 
According to Algorithm \ref{alg:pi} and Eq. \eqref{eq:hpp}, the $\frp_u$ returned at Line 21 in Algorithm \ref{alg:fwd} satisfies
\begin{align*}
\frp_u(u_i)&=\textstyle\sum_{u_j\in U}\sum_{\ell=0}^{t}{\fresidue_u(u_j)\cdot\alpha(1-\alpha)^{\ell}\cdot\left(\UM\cdot\VM\right)^{\ell}(u_j,u_i)}\\
&\textstyle=\sum_{u_j\in U}\sum_{\ell=0}^{\infty}{\fresidue_u(u_j)\cdot\alpha(1-\alpha)^{\ell}\cdot \PM^{\ell}(u_j,u_i)}\\
&\textstyle\quad-\sum_{u_j\in U}\sum_{\ell=t+1}^{\infty}{\fresidue_u(u_j)\cdot\alpha(1-\alpha)^{\ell}\cdot\PM^{\ell}(u_j,u_i)}\\
&\textstyle\ge \sum_{u_j\in U}{\fresidue_u(u_j)\pi(u_j,u_i)}-\sum_{u_j\in U}\sum_{\ell=t+1}^{\infty}{\fresidue_u(u_j)\alpha(1-\alpha)^{\ell}}
\end{align*}
For convenience, let $\fppr^{\prime}(u,u_i)$ denote the approximate HPP obtained at Line 22. Hence,
\begin{align*}
\fppr(u,u_i)&\textstyle =\fppr^{\prime}(u,u_i)+\frp_u(u_i)\\
&\textstyle\ge \fppr^{\prime}(u,u_i)+\sum_{u_j\in U}{\fresidue_u(u_j)\pi(u_j,u_i)} \\
&\textstyle\quad -\sum_{u_j\in U}\sum_{\ell=t+1}^{\infty}{\fresidue_u(u_j)\alpha(1-\alpha)^{\ell}}\\
&\textstyle=\ppr(u,u_i)-\sum_{u_j\in U}\sum_{\ell=t+1}^{\infty}{\fresidue_u(u_j)\cdot\alpha(1-\alpha)^{\ell}}
\end{align*}
Since $t=\log_{\frac{1}{1-\alpha}}{\frac{\sum_{u_i\in U}{\fresidue_u(u_i)}}{\epsilon_f}}-1$, we have $$\textstyle\sum_{u_j\in U}\sum_{\ell=t+1}^{\infty}{\fresidue_u(u_j)\cdot\alpha(1-\alpha)^{\ell}}\le \epsilon_f,$$
which yields $\textstyle\fppr(u,u_i) \ge \ppr(u,u_i)-\epsilon_f$ and completes our proof.
\end{proof}

\begin{proof}[\bf Proof of Lemma \ref{lem:lambda}]
According to Algorithm \ref{alg:pi}, we have
\begin{equation}\label{eq:max-pi}
\textstyle \boldsymbol{\rho}=\mathbf{1}\cdot \sum_{\ell=0}^{\tau}{\alpha(1-\alpha)^{\ell}\cdot\PM^{\ell}},
\end{equation}
when inputting $\mathbf{e}=\mathbf{1}$ and $t=\tau$.
Note that
\begin{align*}
\textstyle \sum_{u_j\in U}{\pi(u_j,u_i)}& \textstyle=\boldsymbol{\rho}(u_i)+\left(\mathbf{1}\cdot \sum_{\ell=\tau+1}^{\infty}{\alpha(1-\alpha)^{\ell}\cdot\PM^{\ell}}\right)(u_i)\\
&\textstyle \le \boldsymbol{\rho}(u_i)+|U|\cdot \left(1-\sum_{\ell=0}^{\tau}{\alpha(1-\alpha)^{\ell}}\right)\\
& = \boldsymbol{\rho}(u_i)+|U|\cdot (1-\alpha)^{\tau+1},
\end{align*}
implying that $\max_{u_i\in U}{\sum_{u_j\in U}{\pi(u_j,u_i)}}\le \max_{u_i\in U}{\boldsymbol{\rho}(u_i)}+|U|\cdot (1-\alpha)^{\tau+1}$. Besides, according to Lemma \ref{lem:asym},
\begin{align*}
\textstyle \sum_{u_j\in U}{\pi(u_j,u_i)}&\textstyle=\sum_{u_j\in U}{\pi(u_i,u_j)\cdot \frac{ws(u_i)}{ws(u_j)}},
\end{align*}
leading to $\max_{u_i\in U}{\sum_{u_j\in U}{\pi(u_j,u_i)}}\le \frac{\max_{u_i\in U}{ws(u_i)}}{\min_{u_j\in U}{ws(u_j)}}$.  This completes our proof.
\end{proof}

\begin{proof}[\bf Proof of Theorem \ref{lem:main}]
Since $\epsilon_f=\epsilon-\epsilon_b$, combining Theorems \ref{lem:fwd} and \ref{lem:bwd} proves the theorem.
\end{proof}

\end{document}